\documentclass[twocolumn,letterpaper,aps,prb,showpacs,floatfix]{revtex4}
\usepackage{graphicx}    % for including graphics
\usepackage{hyperref}    % links
\usepackage{bm}          % bold math letters
\usepackage[sumlimits,intlimits]{amsmath}
\usepackage{amsfonts,amssymb}
\usepackage{multirow}

\begin{document}

\title{Determination of the electronic structure of bilayer graphene
from infrared spectroscopy results}
\author{L. M. Zhang}
% \email{mlzhang@ucsd.edu}
\author{Z. Q. Li}
\author{D. N. Basov}
\author{M. M. Fogler}
\affiliation{University of California San Diego, 9500 Gilman Drive, La
Jolla, California 92093}
\author{Z. Hao and M. C. Martin}
\affiliation{Advanced Light Source Division, Lawrence Berkeley National Laboratory,
Berkeley, California 94720}

\date{\today}

\begin{abstract}

We present an experimental study of the infrared conductivity, transmission, and
reflection of a gated bilayer graphene and their theoretical analysis within the
Slonczewski-Weiss-McClure (SWMc) model. The infrared response is shown to be
governed by the interplay of the interband and the intraband transitions among
the four bands of the bilayer. The position of the main conductivity peak at the
charge neutrality point is determined by the interlayer tunneling frequency. The
shift of this peak as a function of the gate voltage gives information about
less known parameters of the SWMc model, in particular, those responsible for
the electron-hole and sublattice asymmetries. These parameter values are shown
to be consistent with recent electronic structure calculations for the bilayer
graphene and the SWMc parameters commonly used for the bulk graphite.

\end{abstract}


\maketitle

%%%%%%%%%%%%%%%%%%%%%%%%%%%%%%%%%%%%%%%%%%%%%%%%%%%%%%%%%%%%%%%%%%%%%%%%%%%%%%%%
\section{Introduction} \label{sec:Introduction}

Since the monolayer graphene was isolated~\cite{Novoselov2004} and shown to
exhibit the quantum Hall effect~\cite{Novoselov2005tdg, Zhang2005eoq} a few
years ago, ultrathin carbon systems have attracted tremendous
attention.~\cite{graphene_review} Their electron properties are quite unique.
Monolayer graphene has a vanishing Fermi point at the Brillouin zone corner and
low energy quasiparticles with a linear spectrum, $\varepsilon(\mathbf{k}) = \pm
v |\mathbf{k}|$, which obey a massless Dirac equation. Here $\mathbf{k}$ is the
deviation of the crystal momentum from the Brillouin zone corner ($K$ point), $v
= (3 / 2) \gamma_0 a / \hbar$ is the quasiparticle velocity, $\gamma_0$ is the
nearest-neighbor hopping parameter, and $a = 1.42\,\text{\AA}$ is the
carbon-carbon distance. Graphene is the basic building block of other types of
carbon materials. Indeed, the first calculation of its band structure by
Wallace~\cite{Wallace1947} was motivated by his studies of graphite. Extending
that work, Slonczewski and Weiss,~\cite{Slonczewski1958}
McClure,~\cite{McClure1957,Slonczewski1958} and others~\cite{Carter1953} have
developed the now commonly used Slonczewski-Weiss-McClure (SWMc) model for the
low-energy electron properties of graphite. This model, which is equivalent to a
tight-binding model with seven parameters,~\cite{Partoens2006} has proven to be
a very useful analytical tool. It permitted theoretical calculations of a vast
number of properties of graphite, including its diamagnetic susceptibility,
de~Haas-van~Alfven effect, magnetooptical response, cyclotron resonance, and so
on. These properties were actively studied experimentally until the late 70's
and lead to accurate estimates of the principal SWMc parameters, $\gamma_0$
through $\gamma_3$. Still, it proved challenging to unambigously determine the
remaining three SWMc constants $\gamma_4$, $\gamma_5$, and $\Delta$, which are
measured in tens of meV. For illustration, in Table~\ref{tbl:SWMc_values} we
list inequivalent parameter sets from the latest original sources,
Refs.~\onlinecite{Dresselhaus_graphite_review} and \onlinecite{Dillon1977}.
Subsequently, the issue was further confounded by numerous misprints in
reference books and reviews.~\cite{comment_on_typo}. The density-functional
theory calculations,~\cite{Tatar1982,Charlier1991,Gruneis2008} which normally
have accuracy of $\sim 0.1\,\text{eV}$ for quasiparticle dispersion, have not
yet settled this discrepancy.

%%%%%%%%%%%%%%%%%%%%%%%%%%%%%%%%%%%%%%%%%%%%%%%%%%%%%%%%%%%%%%%%%%%%%%%%%%
%
% FIG. 1: Band structure and possible transitions
%
\begin{figure}[b!]
  \includegraphics[width=0.60\columnwidth]{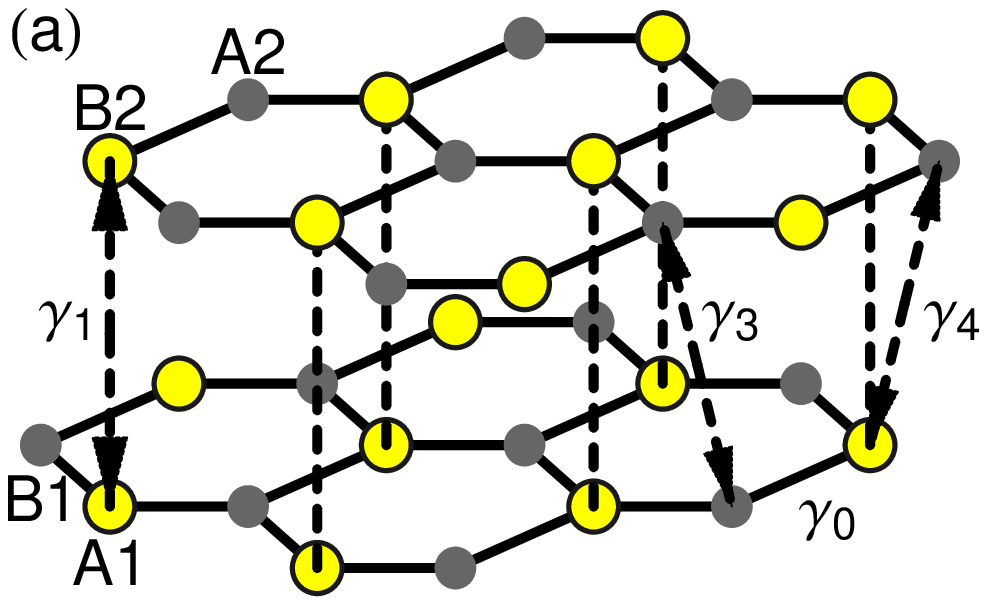}\\
  \includegraphics[width=0.45\columnwidth]{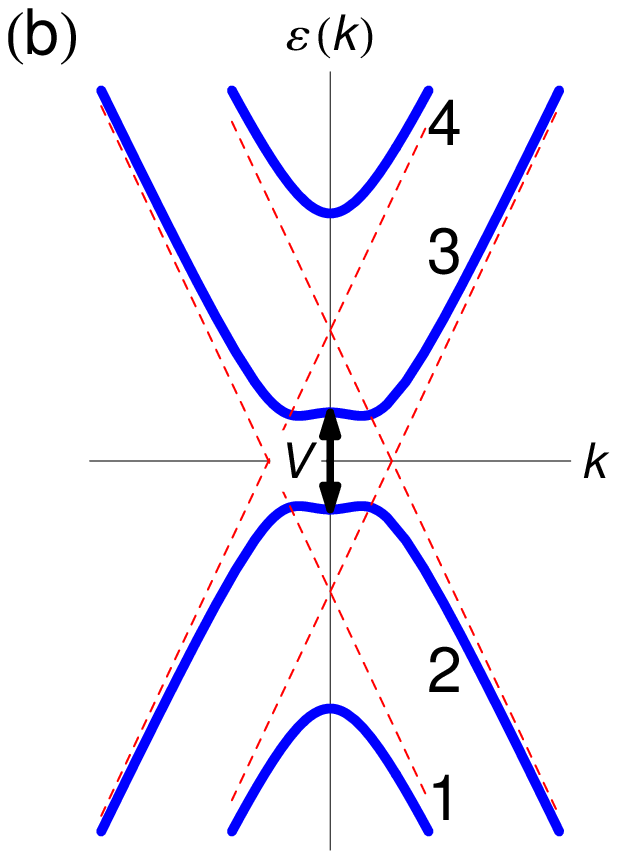}
  \includegraphics[width=0.45\columnwidth]{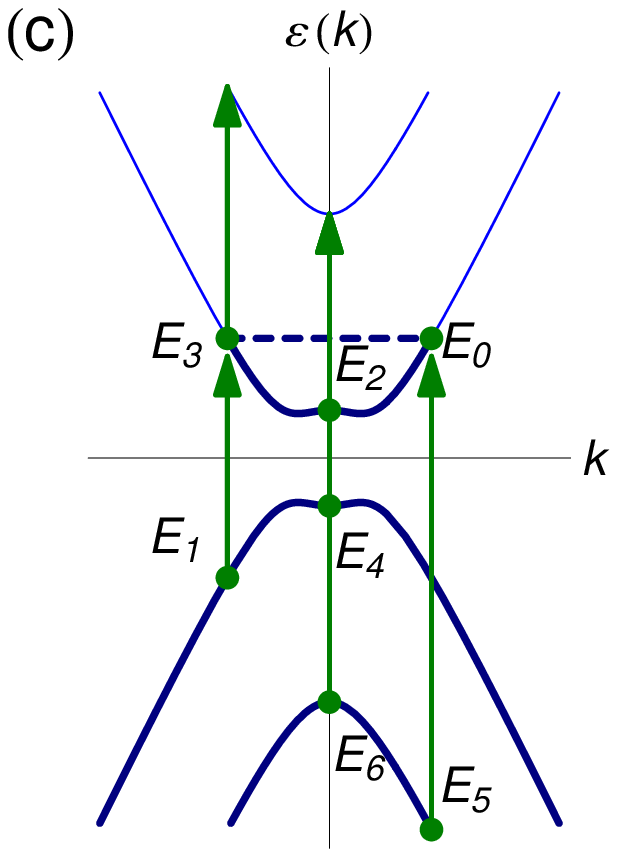}
  \caption{\label{fig:band} (Color online) (a) Crystal structure of the
  graphene bilayer with the relevant SWMc hopping parameters shown. (b) Band
  structure of a biased bilayer (lines), which can be considered as
  hybridization of two shifted Dirac cones (dots). Numbers on the right label
  the four bands. (c) Examples of the allowed optical transitions for the
  chemical potential indicated by the dashed line. Occupied states are shown by
  the thicker lines. The dots and the arrows mark the initial and the final
  states, respectively, of the transitions that produce features at frequencies
  $E_j$, $j = 1, 2, \ldots, 6$ in Fig.~\ref{fig:conductivity}(a) below. $E_0$
  is the intraband transition (Drude peak).}
\end{figure}
%%%%%%%%%%%%%%%%%%%%%%%%%%%%%%%%%%%%%%%%%%%%%%%%%%%%%%%%%%%%%%%%%%%%%%%%%%

In view of the reinvigorated interest to graphene, it has become an important
question to obtain the SWMc constants for a few layer graphene and also to
compare them with those for bulk graphite. Thus, some difference
between the graphite and a graphene bilayer was recently reported, based on the
analysis of Raman scattering.~\cite{Malard2007} Several \textit{ab initio\/}
calculations of these parameters for the bilayer have also been
done.~\cite{Trickey1992,Latil2006,Latil2007,Min2007,Aoki2007} Unfortunately,
they have not explicitly discussed the less accurately known SWMc parameters.

%%%%%%%%%%%%%%%%%%%%%%%%%%%%%%%%%%%%%%%%%%%%%%%%%%%%%%%%%%%%%%%%%%%%%%%%%%
%
% Table 1
%
\begin{table*}
\label{tbl:SWMc_values}
\caption{The SWMc parameters (in eV) according to previous and present work. The numbers in
parentheses are the reported accuracy of the trailing
decimals. The ``Exp'' and ``DFT'' stand for experiment and density functional theory, respectively.}
\begin{tabular*}{1.00\textwidth}%
      {c@{\hspace{0.4in}}r@{.}lr@{.}lr@{.}l%
        @{\hspace{0.6in}}r@{.}lr@{.}lr@{.}lr@{.}l%
        @{\hspace{0.4in}}r@{.}lr@{.}l}
\hline\hline
SWMc      & \multicolumn6c{Graphene bilayer} & \multicolumn8c{Graphite, early work}  & \multicolumn4c{Graphite, recent work}\\
parameter & Pres & \,work
          & \multicolumn2l{Exp\footnotemark[1]}
          & \multicolumn2l{DFT\footnotemark[2]}
          & \multicolumn2l{Exp\footnotemark[3]}
          & \multicolumn2l{Exp\footnotemark[4]}
          & \multicolumn2l{DFT\footnotemark[5]}
          & \multicolumn2l{DFT\footnotemark[6]}
          & \multicolumn2l{Exp\footnotemark[7]}
          & \multicolumn2l{DFT\footnotemark[8]} \\
\hline
$\gamma_0$& 3&0\footnotemark[9]     & 2&9      & 2&6      & 3&16(5)  & 3&11     & 2&92      & 2&598(15) & \multicolumn2c{} & \multicolumn2c{}\\
$\gamma_1$& 0&40(1)                 & 0&30     & 0&3      & 0&39(1)  & 0&392    & 0&27    & 0&364(20) & \multicolumn2c{} & \multicolumn2c{}\\
$\gamma_2$& 0&0\footnotemark[10]
                                    & 0&0\footnotemark[10]
                                               & 0&0\footnotemark[10]
                                                          &-0&020(2) &-0&0201   &-0&022   &-0&014(8)  & \multicolumn2c{} & \multicolumn2c{}\\
$\gamma_3$& 0&3\footnotemark[9]
                                    & 0&10     & 0&3      & 0&315(15)& 0&29     & 0&15    & 0&319(20) & \multicolumn2c{} & \multicolumn2c{}\\
$\gamma_4$& 0&15(4)                 & 0&12     & \multicolumn2c{}
                                                          & 0&044(24)& 0&124    & 0&10    & 0&177(25) & \multicolumn2c{} & \multicolumn2c{}\\
$\gamma_5$& 0&0\footnotemark[10]
                                    & 0&0\footnotemark[10]
                                               & 0&0\footnotemark[10]
                                                          & 0&038(5) & 0&0234   & 0&0063    & 0&036(13) & \multicolumn2c{} & \multicolumn2c{}\\
$\Delta$  & 0&018(3)                & \multicolumn2c{}
                                               & 0&01\footnotemark[11]
                                                          &-0&008(2) &-0&0049   & 0&0079  &-0&026(10) & $<$0&01\footnotemark[12] & -0&037\footnotemark[13]\\
$\Delta^\prime = \Delta - \gamma_2 + \gamma_5$
          & 0&018(3)                & \multicolumn2c{}
                                               & 0&01\footnotemark[11]
                                                          & 0&037(5) & 0&0386   & 0&0362  & 0&024(18) & \multicolumn2c{} & \multicolumn2c{}\\
\hline\hline
\end{tabular*}
\begin{minipage}{1.00\textwidth}

  \footnotetext[1]{L.~M.~Malard~\textit{et al}\/., 
  Phys.\ Rev.\ B\ \textbf{76}, 201401 (2007).~\cite{Malard2007}}

  \footnotetext[2]{H.~Min, B.~R.~Sahu, S.~K.~Banerjee, and A.~H.~MacDonald,
  Phys.\ Rev.\ B\ \textbf{75}, 155115 (2007).~\cite{Min2007}}

  \footnotetext[3]{M.~S.~Dresselhaus and G.~Dresselhaus,
  Adv.\ Phys.\ \textbf{30}, 139 (1981).~\cite{Dresselhaus_graphite_review}}

  \footnotetext[4]{R.~O.~Dillon, I.~L.~Spain, and J.~W.~McClure,
  J.\ Phys.\ Chem.\ Solids\ \textbf{38}, 635 (1977).~\cite{Dillon1977}}

  \footnotetext[5]{R. C. Tatar and S. Rabii,
  Phys.\ Rev.\ B\ \textbf{25}, 4126 (1982).~\cite{Tatar1982}}

  \footnotetext[6]{J.-C. Charlier, X. Gonze, and J.-P. Michenaud,
  Phys.\ Rev.\ B\ \textbf{43}, 4579 (1991).~\cite{Charlier1991}}

  \footnotetext[7]{M.~Orlita \textit{et al\/}.,
  Phys.\ Rev.\ Lett.\ \textbf{100}, 136403 (2008).~\cite{Orlita2008}}

  \footnotetext[8]{A.~Gr\"uneis \textit{et al\/}.,
  Phys.\ Rev.\ Lett.\ \textbf{100}, 037601 (2008).~\cite{Gruneis2008}}

  \footnotetext[9]{This value cannot be very accurately found from our analysis and is instead adopted
                   from the literature.}
  \footnotetext[10]{Physically irrelevant in the bilayer but should be set to zero for
                    calculating $\Delta^\prime$ from $\Delta$.}

  \footnotetext[11]{Our estimate based on digitizing band dispersion graphs published in
  Refs.~\onlinecite{Min2007,Latil2006,Latil2007,Aoki2007}.}

  \footnotetext[12]{Absolute value only.}

  \footnotetext[13]{The negative sign (omitted in Ref.~\onlinecite{Gruneis2008}) is required
  for consistency with the conventional definition~\cite{McClure1957} of $\Delta$.}

\end{minipage}
\end{table*}
%%%%%%%%%%%%%%%%%%%%%%%%%%%%%%%%%%%%%%%%%%%%%%%%%%%%%%%%%%%%%%%%%%%%%%%%%%

The bilayer is a system intermediate between graphene and bulk graphite. Its
lattice structure (for the case of the Bernal or AB stacking) is illustrated in
Fig.~\ref{fig:band}(a). The corresponding band
structure,~\cite{McCann2006a, McCann2007, Nilsson2008} shown in
Fig.~\ref{fig:band}(b), consists of four bands. These bands arise from splitting
and hybridization of the Dirac cones of the individual layers by the interlayer
hopping matrix element $\gamma_1$ and by the electrostatic potential difference
$V$ between the two layers.~\cite{McCann2006a, Guinea2006} The latter can be
controlled experimentally by varying the voltage $V_g$ of a nearby metallic
gate~\cite{Castro2007, Oostinga2007} or by doping.~\cite{Ohta2006} This degree
of tunability makes the bilayer graphene an extremely interesting material for
both fundamental study and applications.

In this paper we show that $\gamma_1$, $v_4\equiv\gamma_4/\gamma_0$, and
$\Delta$ can be \emph{directly} extracted from the dynamical
condictivity measured in zero magnetic field. This is in contrast to the bulk
graphite where determination of the SWMc constants was never straightforward
and almost invariably required the use of strong magnetic fields.

The dynamical conductivity $\sigma(\Omega)$ is determined by the six possible
transitions among the four bands, see Fig.~\ref{fig:band}(c). They have energies
of the order of a few $10^{-1}\,\text{eV}$, which is in the infrared optical
range. Recently, experimental measurements of the infrared response of the
bilayers have been carried out by our~\cite{Li2008} and other~\cite{Wang2008,
Kuzmenko2008} groups. Below we identify and explain the key findings of these
experiments based on how different combinations of the interband transitions are
either activated or suppressed by the Pauli exclusion principle. Our theory
enables us to reach a quantitative agreement with the experiment using SWMc
$\gamma_0$, $\gamma_1$, $\gamma_4$, and $\Delta$, and also the phenomenological
broadening constant $\Gamma$ as adjustable parameters. The values of the SWMc
parameters that give the best fit are given in the second column of
Table~\ref{tbl:SWMc_values}. Note that the next-nearest layer hopping parameters
$\gamma_2$ and $\gamma_5$ are irrelevant for the bilayer. The parameter
$\gamma_3$ cannot be reliably estimated from these particular experiments
because it has an effect similar to the simple broadening ($\Gamma$) in the
range of carrier concentrations suitable for our analysis.

\begin{figure}
  \centering
  \includegraphics[width=0.75\columnwidth]{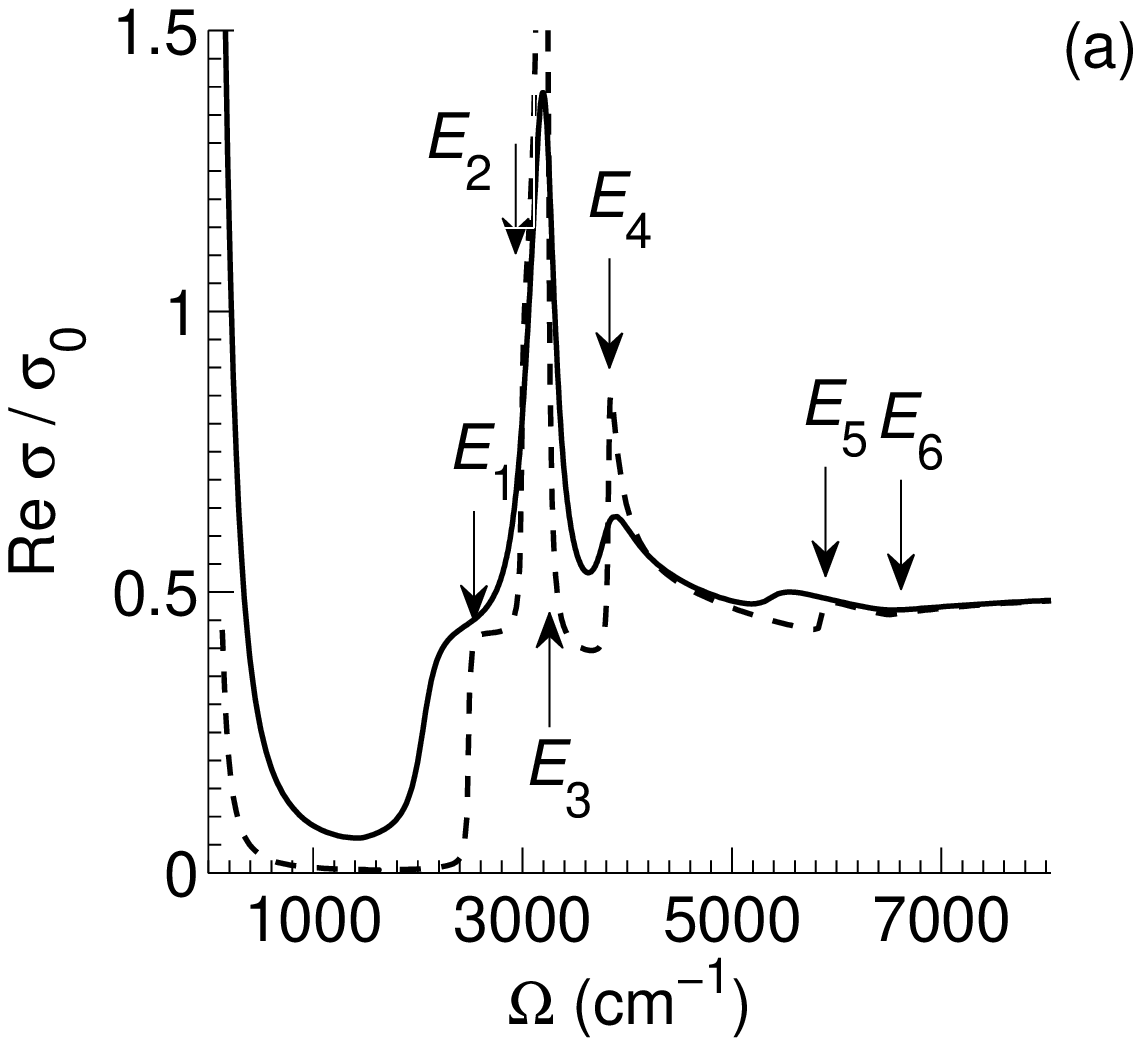}
  \includegraphics[width=0.75\columnwidth]{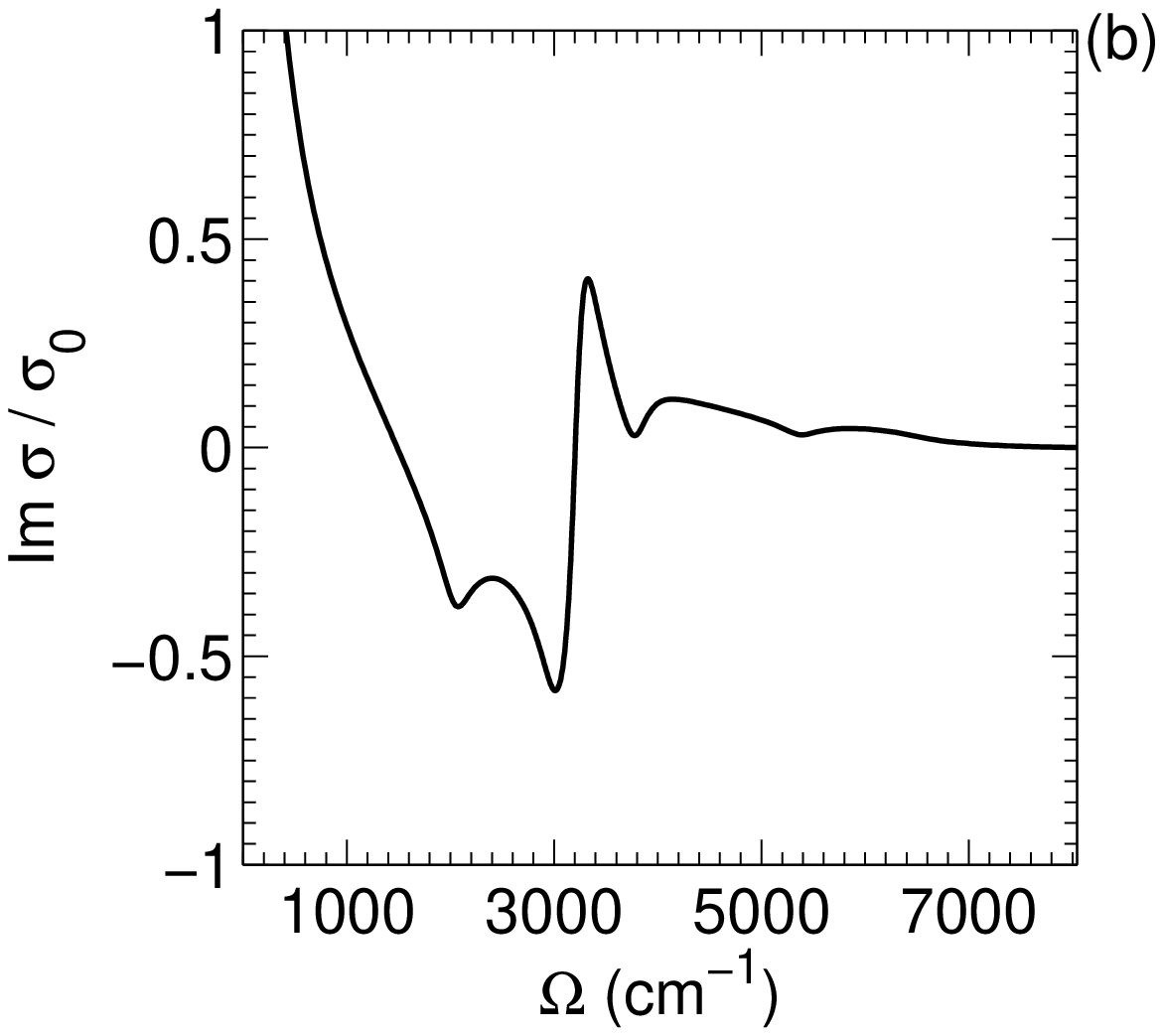}
\caption{(a) Real and (b) imaginary part of conductivity
in units of $\sigma_0 = {e^2}/{\hbar}$ for the gate voltage $\delta V =
-100\,\text{V}$. The solid curves are for
broadening $\Gamma = 0.02\gamma_1$. The dashed curve is for $\Gamma =
0.002 \gamma_1$. 
  \label{fig:conductivity}}
\end{figure}

Previous theoretical studies of the optical conductivity of bilayer
graphene~\cite{McCann2007, Abergel2007, Nilsson2008, Benfatto2008, Nicol2008}
used a simplified model in which only $\gamma_0$ and $\gamma_1$ were taken into
account. This model successfully explains the major features of
$\sigma(\Omega)$ as well as its dependence on the gate voltage $V_g$, and we
qualitatively summarize it as follows. Conduction and valence bands are
symmetric. In the absence of the electrostatic potential difference $V$ between
the layers the two conduction (valence) bands have the same shape and are
shifted by $\gamma_1$.  Except the range of very small momenta $k$, their shape
remain nearly identical even in the presence of a finite $V$. As a result,
there is a high optical density of states for transitions between the two pairs
of bands at frequency $\gamma_1 / \hbar$, which gives rise to a sharp peak in
the real part of the conductivity $\text{Re}\,\sigma(\Omega)$ at $\Omega =
\gamma_1 / \hbar \approx 3200\,\text{cm}^{-1}$ (using $\gamma_1 =
0.40\,\text{eV}$). Other transitions give more gradually varying contributions
to $\text{Re}\,\sigma(\Omega)$, eventually leading to the asymptotic
``universal'' value~\cite{Ando2002, McCann2007, Abergel2007, Nilsson2008, Benfatto2008,
Nicol2008, Nair2008} $\sigma = e^2 / 2\hbar$ at high frequency (which is twice the value
for the monolayer~\cite{Gusynin2007}).  Finally, in real graphene systems the
conductivity features are never sharp because of a finite lifetime due to,
e.g., disorder scattering. This broadens the peaks and can also merge together
several features that are close in energy, see Fig.~\ref{fig:conductivity}.

%%%%%%%%%%%%%%%%%%%%%%%%%%%%%%%%%%%%%%%%%%%%%%%%%%%%%%%%%%%%%%%%%%%%%%%%%
%
% FIG. 2
%
%%%%%%%%%%%%%%%%%%%%%%%%%%%%%%%%%%%%%%%%%%%%%%%%%%%%%%%%%%%%%%%%%%%%%%%%%

Our recent infrared experiments~\cite{Li2008} as well as
measurements by another group~\cite{Kuzmenko2008} have largely
confirmed this picture but also found features that cannot be explained
within this simple model. In particular, the conductivity peaks on the
electron and the hole sides are displaced in energy from $\gamma_1$ by
about $10\%$ in opposite directions. [Electron and hole doping
is identified with, respectively, positive and negative $\delta V = V_g
- V_\text{CN}$, where $V_\text{CN}$ is the gate voltage at which the
bilayer is tuned to the charge-neutrality (CN) point.]

In order to investigate the origin of these features in this paper we
carry out a combined experimental-theoretical study of the infrared
response of a bilayer graphene. We attribute the observed electron-hole
asymmetry to the effect of $\gamma_4$ and $\Delta$. We find that
including these parameters is essential for a more accurate discussion
of $\sigma(\Omega)$ of the bilayer. Besides differences in the optical
response, $\gamma_4$ and $\Delta$ also make effective masses for
electrons and holes unequal,~\cite{comment_on_g7} in agreement with the
findings from the Raman scattering.~\cite{Malard2007}

In our experiments, we have measured the optical reflection $R(\Omega, V_g)$
and transmission $T(\Omega, V_g)$ as a function of the frequency
$\Omega$ and the gate voltage $V_g$. From $R$ and $T$ we extracted the
real and imaginary part of the conductivity using a commercial software
package. Some of these experimental results were reported
previously.~\cite{Li2008}

In this paper we present more extensive experimental data and we also compute
the same three quantities --- $\sigma$, $R$, and $T$ --- theoretically. The
calculation requires accounting for the interplay of several physical phenomena:
(a) electrostatic charging of the layers (b) their dynamical conductivity, (c)
disorder, and (d) the optical properties of the environment (sample, substrate,
and the gate). Each of these ingredients has been studied in the
past.~\cite{McCann2006, McCann2007, Abergel2007, Nilsson2008, Benfatto2008,
Nicol2008, Castro2007, Castro2008} Here we carry out all these calculations in a
single paper albeit we include disorder broadening in a very simple way. This
enables us to directly compare our theoretical results with the measurements.

The remainder of the paper is organized as follows. In Sec.~\ref{sec:Results} we
summarize our results. Theoretical derivation is outlined in
Sec.~\ref{sec:Derivation}. Section~\ref{sec:Discussion} contains comparison of
the theory and experiment, discussion, and conclusions. Some calculational
details are relegated to the Appendix.

%%%%%%%%%%%%%%%%%%%%%%%%%%%%%%%%%%%%%%%%%%%%%%%%%%%%%%%%%%%%%%%%%%%%%%%%%%%%%%%%
\section{Results}\label{sec:Results}

\begin{figure}[htb]
  \centering
  \includegraphics[width=0.47\textwidth]{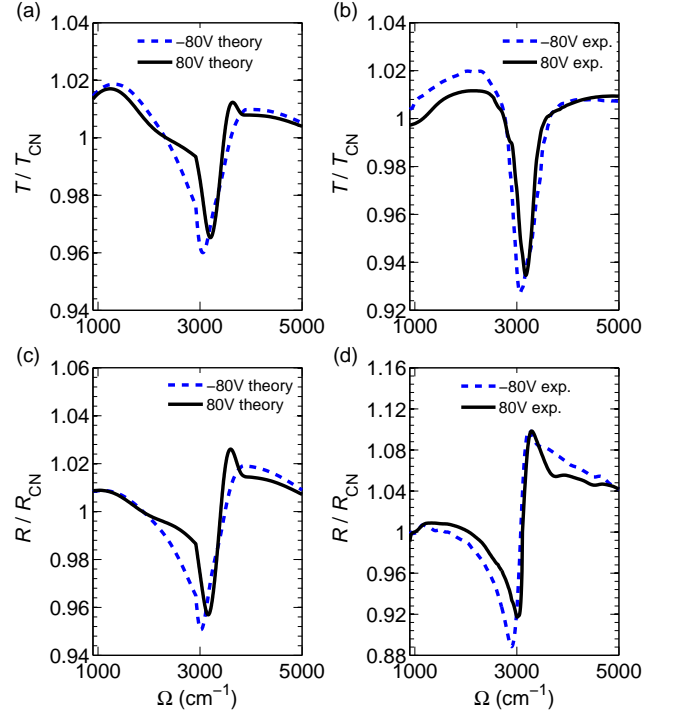}
\caption{\label{fig:bilayeroptics}(Color online)
Relative transmission: (a) theory (b) experiment. Relative
reflection: (c) theory (d) experiment.
The solid line is for electrons, $\delta V \approx
+80\,\text{V}$. The dashed line is for holes, $\delta V \approx
-80\,\text{V}$. The experimental uncertainties are $\sim 0.002$ ($0.2\%$) at
$\Omega$ near $3000\,\text{cm}^{-1}$ and $\sim 0.5\%$ at high frequency.
}
\end{figure}

To measure the optical response of the bilayer we employed synchrotron
infrared radiation, as described previously.~\cite{Li2008a,Li2008}
Understandably, the two-atom thick sample has a rather small optical
signal. The quantity which can be extracted most reliably from the
current experiments is the relative transmission $T(\Omega, V_g) /
T(\Omega, V_\text{CN})$ and reflection $R(\Omega, V_g)/ R(\Omega,
V_\text{CN})$. All measurements were done at the temperature of
$45\,\text{K}$. The data for the largest $|\delta V| = |V_g -
V_\text{CN}|$ are depicted in Fig.~\ref{fig:bilayeroptics}. The main
feature in the relative transmission spectra is a small but clearly
visible dip around $\Omega = 3200\,\text{cm}^{-1}$. Away from the dip,
the relative transmission is slightly higher than unity. The relative
reflection spectra are characterized by a dip-peak structure.
Transmission and reflection spectra are asymmetic between positive and
negative $\delta V$, which correspond, respectively, to doping of
electrons and holes in bilayer graphene.

\begin{figure*}[htb]
  \begin{center}
    \includegraphics[width=0.30\textwidth]{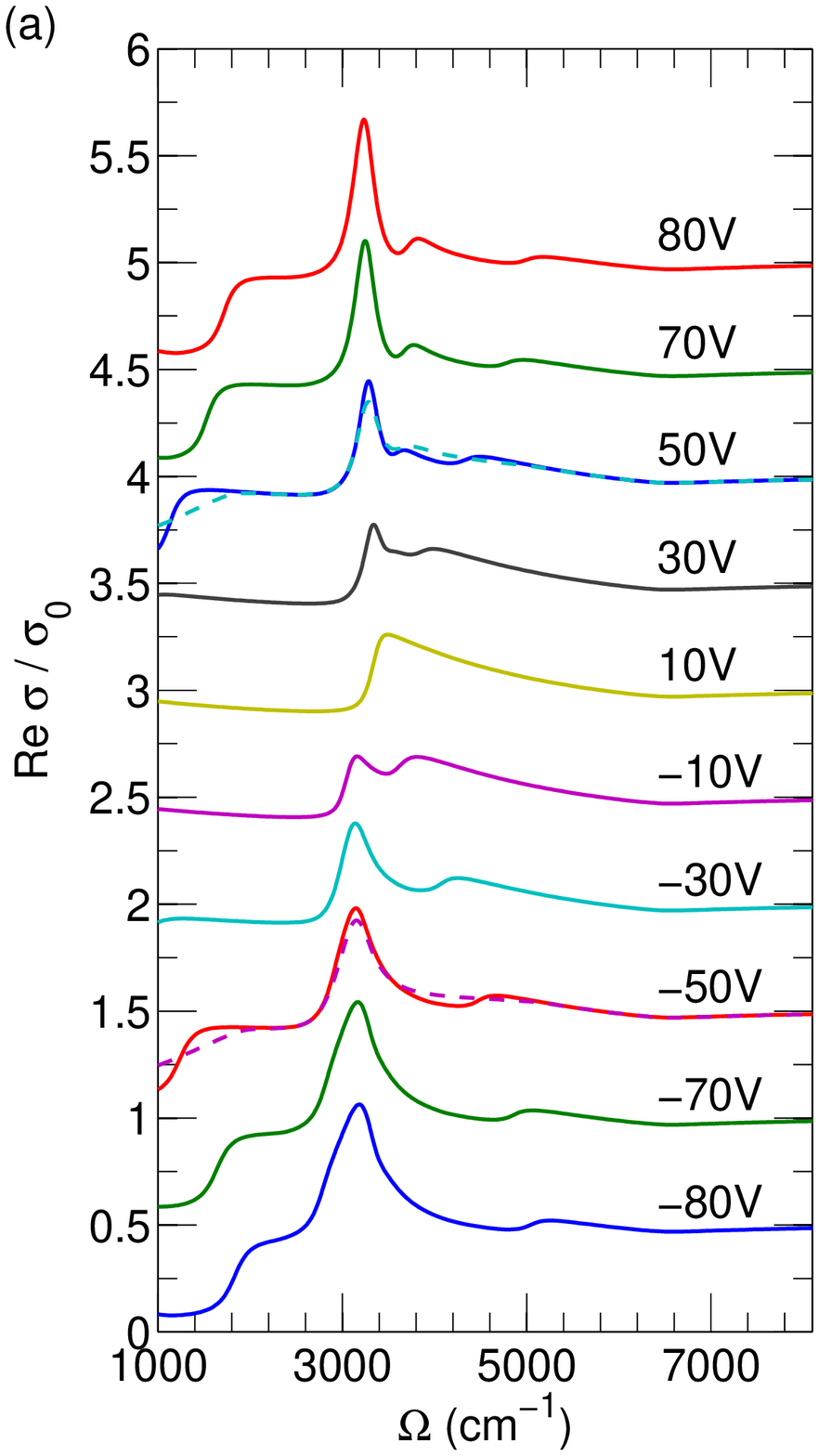}
    \includegraphics[width=0.30\textwidth]{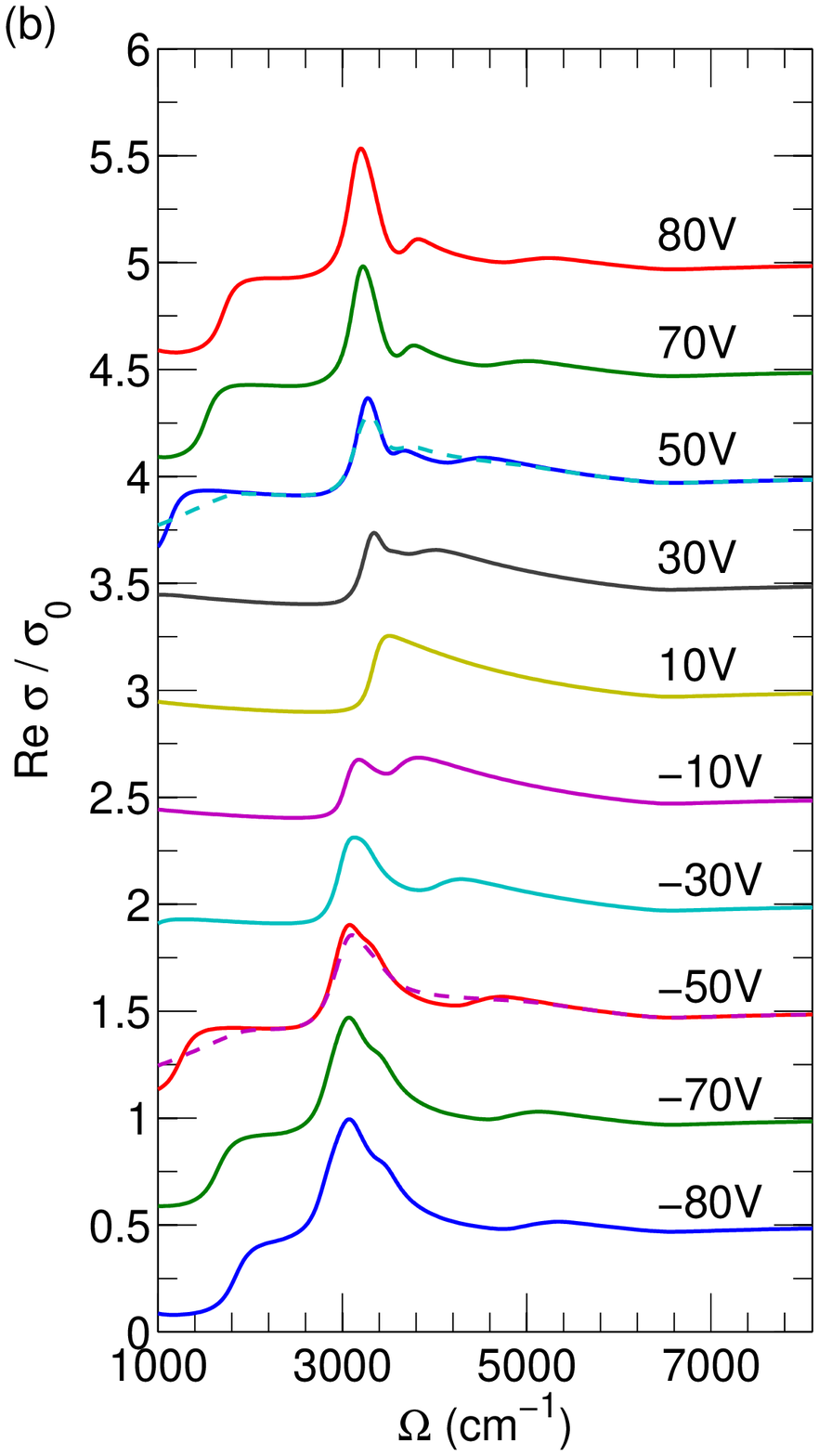}
    \includegraphics[width=0.30\textwidth]{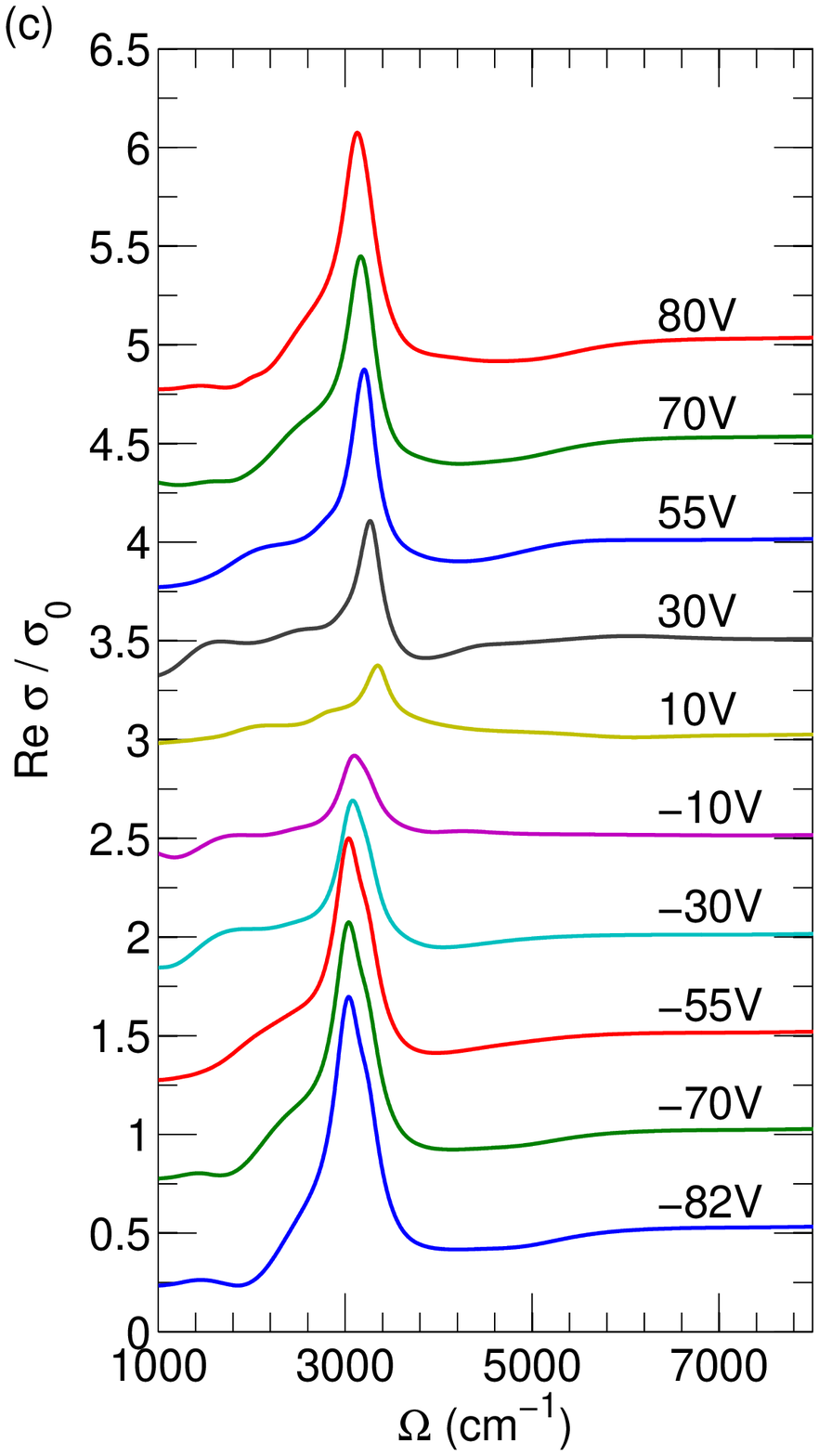}
  \end{center}
  \caption{\label{fig:sig_Omega}(Color online) (a), 
  (b) Theoretical and (c) experimental results for the conductivity $\text{Re}\,\sigma$, in
  units of $\sigma_0 = 4 e^2 / \hbar$, as a function of frequency $\Omega$.
  The deviation $\delta V$ of the gate voltage from the charge neutrality point
  is indicated next to each curve. For clarity, the
  curves are offset vertically by $0.5 \sigma_0$ from one another.
  The SWMc parameters for plot (b) are given in Table~\ref{tbl:SWMc_values}.
  In (a) they are the same, except $\gamma_3$ is set to zero.
  The dashed curves superimposed on the $\delta V = +50\,\text{V}$ ($-50\,\text{V}$) traces in
  (a) and (b) are the arithmetic means of all the positive (negative) $\delta V$
  curves. Their significance is discussed in Sec.~\ref{sec:Discussion}.
  The estimated uncertainty of the measured $\text{Re}\,\sigma$ is $0.125 \sigma_0$
  at $\Omega \sim 8000\,\text{cm}^{-1}$ and $0.0625 \sigma_0$ at $\Omega
  \sim 3000\,\text{cm}^{-1}$.}
  
\end{figure*}

From the transmission and reflection data, we extracted the optical
conductivity.~\cite{Li2007,Li2008a,Li2008} The dominant feature in the
conductivity spectra is a strong peak at $\Omega \approx 3200\,\text{cm}^{-1}$,
see Fig.~\ref{fig:sig_Omega}(c). Below the main peak, we observed a broadened
threshold feature, which shifts systematically with $\delta V$. The most
intriguing observation is again the electron-hole asymmetry in the optical
conductivity. For instance, the frequencies of the main peak in
$\text{Re}\,\sigma(\Omega)$ and its voltage dependence are noticeably different
for electrons and holes, see Fig.~\ref{fig:sig_Omega}(c). Also, while the peak
is quite symmetric at large positive voltages, at high negative $\delta V$, it
is not. The most probable reason is the existence of a secondary peak at a
slightly larger $\Omega$, see below.

On the theory side, we calculated $\sigma$, $T$, and $R$, using the SWMc
constants and $\Gamma$ as adjustable parameters. Results for the conductivity
are shown in Fig.~\ref{fig:sig_Omega}(b). The reflection and transmission are
plotted in Fig.~\ref{fig:bilayeroptics}. The calculational parameters were
adjusted to reproduce the frequency positions and widths of the main features of
the experimental data. Interestingly, in this way of fitting, it was not
possible to achieve an equally good agreement for the vertical scale of the
observed features. Still their qualitative trend as a function of $\delta V$ is
reproduced well.

Both in experiment and in calculations the carrier concentrations are always
smaller than the characteristic value $n_0$ given by
\begin{equation}\label{eq:n_0}
   n_0 = \frac{\gamma_1^2}{\hbar^2 v^2} = 3.7 \times 10^{13} \,\text{cm}^{-
2}\,.
\end{equation}
Here and below we assume that $\gamma_0 = 3.0\,\text{eV}$, which corresponds to
$v = (3/2) \gamma_0 a / \hbar = 1.0 \times 10^8\,\text{cm}/{s}$. (Based on
other results in the literature, this value should be accurate to about
$10\%$.) At concentrations $|n| < n_0$ the high energy bands 1 and 4 have no free
carriers and $\text{Re}\,\sigma(\Omega)$ has a pronounced peak at $\Omega
\approx 3200\,\text{cm}^{-1}$. As explained above, this feature corresponds to
transition between band pairs that are nearly parallel: bands 3 and 4 for $\mu >
0$ or bands 1 and 2 for $\mu<0$, see Fig.~\ref{fig:band}.

The evolution of the infrared response with $V_g$ can be understood as follows.
As the gate voltage deviates further away from $V_\text{CN}$, the electron
concentration
\begin{equation}\label{eqn:n_from_V}
                      n = C_b \delta V / e
\end{equation}
and the chemical potential $\mu$ increase by the absolute value. Here $C_b$ is
the capacitance between the bilayer and the gate. As a result of an increased
$|n|$, the peak become more pronounced. Simultaneously, near the higher
frequency side of the peak a depletion of conductivity develops. One can say
that the optical weight is increasingly transfered from the high frequencies to
the $\gamma_1$ peak. Larger conductivity is directly associated with decreased
optical transmission. Therefore one observes an increasing dip in the
transmission near $\gamma_1$ and a higher transmission at higher $\Omega$, see
Fig.~\ref{fig:bilayeroptics}. Similar features appear in the reflection but they
are more difficult to interpret as they are also affected by
$\text{Im}\,\sigma(\Omega)$.

%%%%%%%%%%%%%%%%%%%%%%%%%%%%%%%%%%%%%%%%%%%%%%%%%%%%%%%%%%%%%%%%%%%%%%%%%%%%%%%%
%
% Fig.3
%
%%
%%%%%%%%%%%%%%%%%%%%%%%%%%%%%%%%%%%%%%%%%%%%%%%%%%%%%%%%%%%%%%%%%%%%%%%%%%%%%%%%

%%%%%%%%%%%%%%%%%%%%%%%%%%%%%%%%%%%%%%%%%%%%%%%%%%%%%%%%%%%%%%%%%%%%%%%%%%%%%%%%
%
% Fig.4
%
%
%%%%%%%%%%%%%%%%%%%%%%%%%%%%%%%%%%%%%%%%%%%%%%%%%%%%%%%%%%%%%%%%%%%%%%%%%%%%%%%%

\begin{figure}[htb]
  \begin{center}
    \includegraphics[width=0.85\columnwidth]{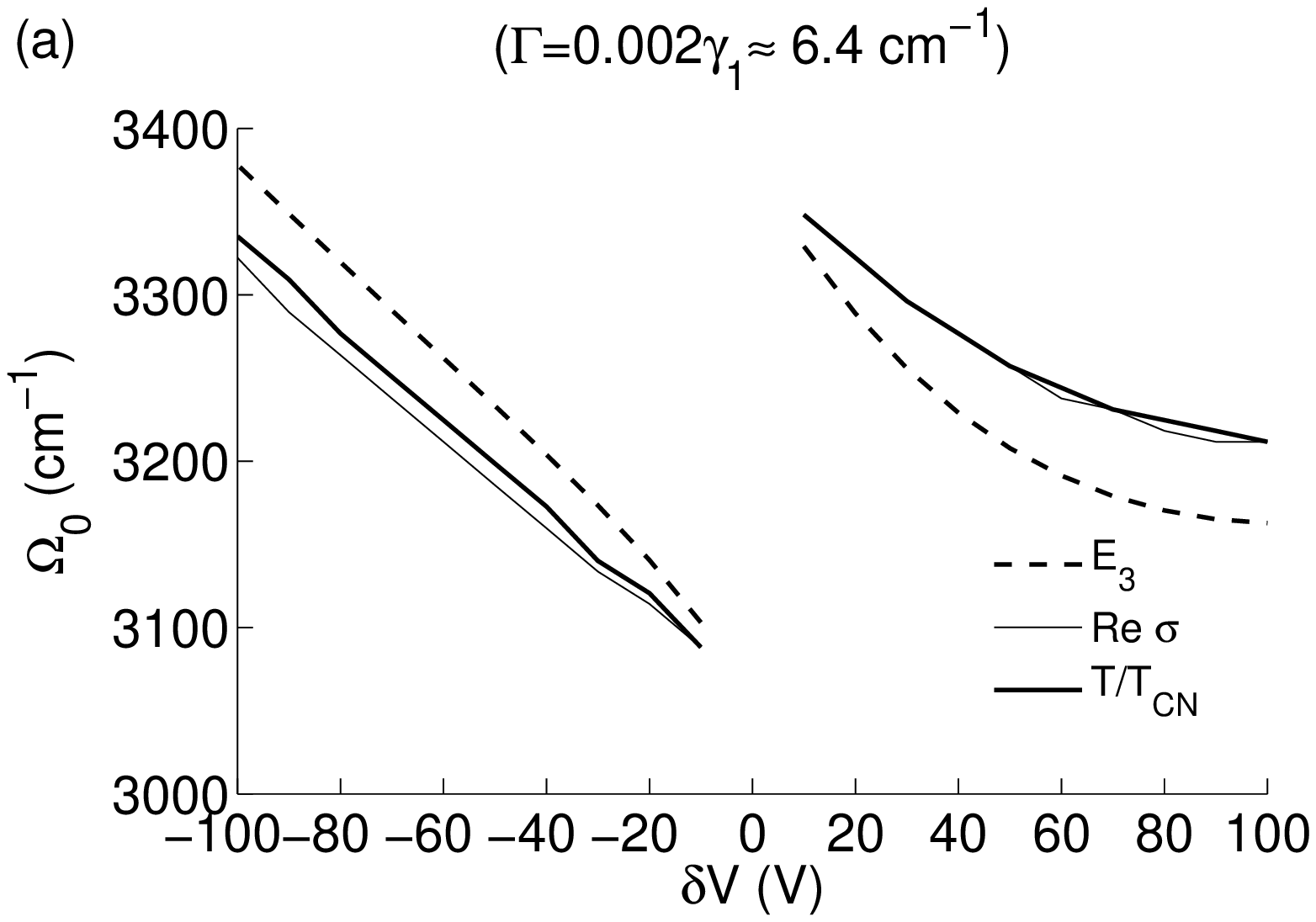}
    \includegraphics[width=0.85\columnwidth]{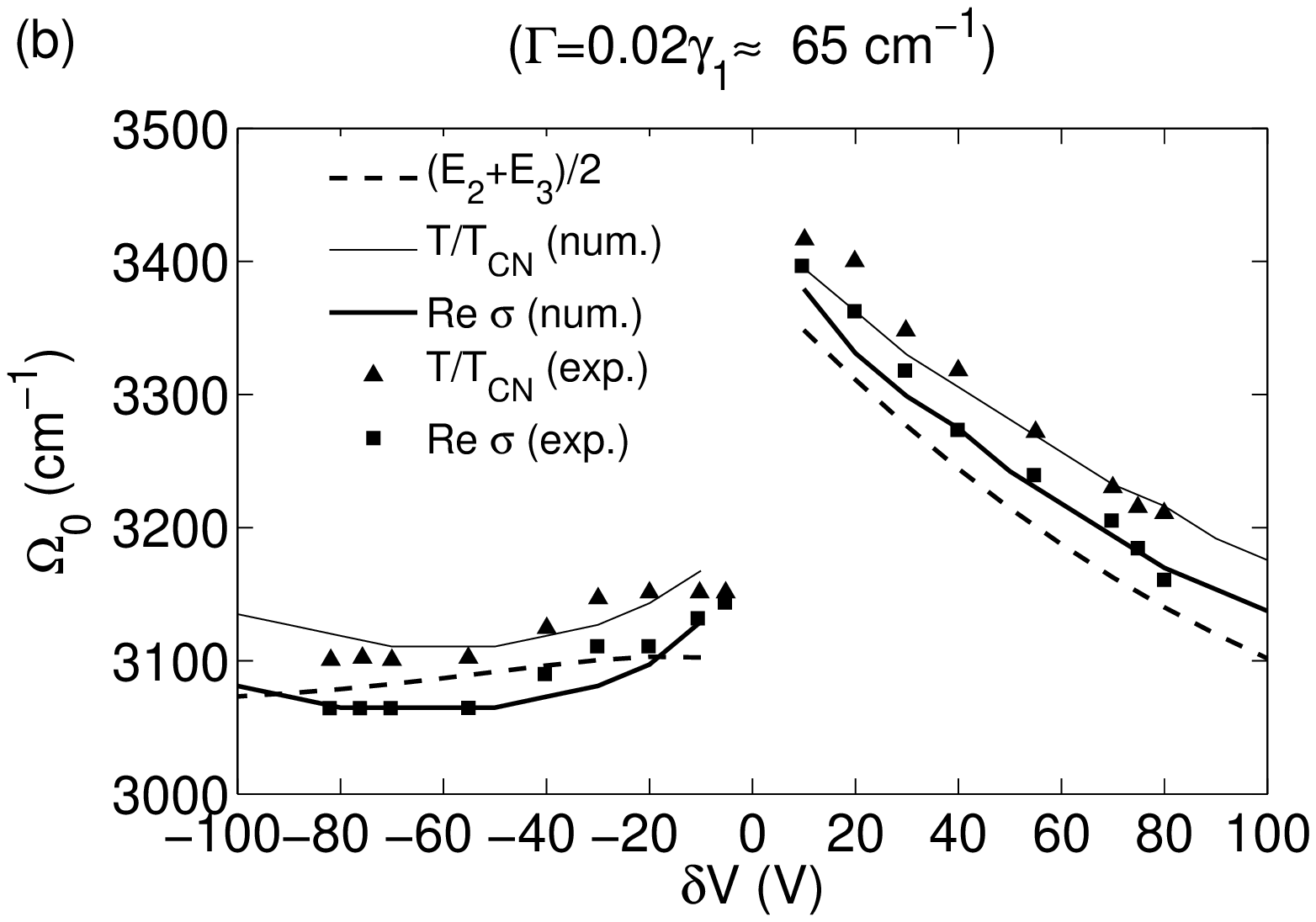}
  \end{center}
\caption{Position of the $\gamma_1$ peak \textit{vs\/}. gate voltage for
the two values of the broadening: (a)~$\Gamma = 0.02\gamma_1$ and
(b)~$\Gamma = 0.002\gamma_1$. The solid lines are our numerical results
from the conductivity; the thick lines are from the relative
transmission. The dashed lines show $E_3$ and $(E_2 + E_3) / 2$ in the
cases (a) and (b), respectively. The SWMc parameters used in the
calculation are listed in the first column of
Table~\ref{tbl:SWMc_values}, except in (a) $\gamma_3$ is set to zero.
The symbols are the peak positions determined from the measured
conductivity (squares) and transmission (triangles).
\label{fig:Tpeaks}}
\end{figure}

Very important for our analysis are the aforementioned small shifts in
the position of the $\gamma_1$ peak as a function of $\delta V$. Within
the SWMc model, their origin is as follows. In the absence of
broadening, the peak arises from the absorption in the range of
frequencies, $E_2 < \hbar \Omega < E_3$, see Figs.~\ref{fig:band} and
\ref{fig:conductivity}. Since the optical weight at $E_3$ is higher, the
conductivity peak occurs at energy $E_3$. However, if the broadening is
large enough, the optical weight becomes distributed more uniformly, and
the peak position moves to the midpoint of $E_2$ and $E_3$, see
Fig.~\ref{fig:Tpeaks}. Energies $E_2$ and $E_3$ themselves vary with the
gate voltage (or $n$). For positive $\delta V$ (positive $n$), $E_2
\equiv E_2^{+}$ is the energy difference between the bands 3 and 4 at $k
= 0$. The energy $E_3 \equiv E_3^{+}$ is the corresponding difference at
$k = k_F$, where
\begin{equation}\label{eq:k_F}
   k_F = \text{sign}\,(n) \sqrt{\pi |n|}
\end{equation}
is the Fermi momentum. For $\delta V < 0$ we denote $E_2$ and $E_3$ by,
respectively, $E_2^{-}$ and $E_3^{-}$ and they are computed using the bands 1
and 2 instead of 3 and 4.

%%%%%%%%%%%%%%%%%%%%%%%%%%%%%%%%%%%%%%%%%%%%%%%%%%%%%%%%%%%%%%%%%%%%%%
%
% FIG. 5
%%
%%%%%%%%%%%%%%%%%%%%%%%%%%%%%%%%%%%%%%%%%%%%%%%%%%%%%%%%%%%%%%%%%%%%%%

From the band structure,~\cite{McCann2007, Nilsson2008} we can find the
following approximate expressions valid for $n \ll n_0$:
\begin{align}
    E^{\pm}_2 &\simeq \gamma_1 - \frac{V}{2} \pm \Delta\,,
 \label{eqn:E2_asym} \\
    E^{\pm}_3 &\simeq \gamma_1 \sqrt{1 + \frac{2 \pi |n|}{n_0} }
          - \sqrt{ \frac{V^2}{4} + \left( \frac{\pi \gamma_1 n}{n_0} \right)^2}
  \notag\\
       &\pm \Delta \mp 2\left( 2 v_4\gamma_1 + \Delta \right) \frac{\pi |n|}{n_0}\,.
  \label{eqn:E3_asym}
\end{align}
Here $V = V(n)$ as well as the chemical potential $\mu = \mu(n)$
are determined self-consistently by the electrostatics of the
system,~\cite{McCann2006} see Sec.~\ref{sec:Derivation}.
These equations indicate that the parameters primarily
responsible for electron-hole asymmetry are $\gamma_4$ and $\Delta$.

\begin{figure}[htb]
  \begin{center}
    \includegraphics[width=0.45\columnwidth]{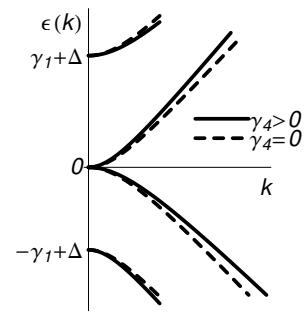}
  \end{center}
  \caption{ The effect of $\gamma_4$ and $\Delta$ on the band structure.
  Parameter $\Delta$ raises the bands 1 and 4. The interlayer neighbor hopping
  term $\gamma_4$ gives a contribution quadratic in $k$ opposite in sign for
  the conduction and the valence bands. The solid (dashed) lines are the bands
  with positive (zero) value of $\gamma_4$.  \label{fig:BandAsym}}
\end{figure}

Parameter $\Delta$ is the difference of the on-site electron energies
of the A and the B sites~\cite{Carter1953, Partoens2006} [the stacked and
unstacked sublattices, respectively, see Fig.~\ref{fig:band}(a)]. It has
two effects: first, it lifts the $k = 0$ energy for bands 1 and 4;
second, it adds a $k$ dependent perturbation to the two band dispersion.
Parameter $v_4 = v \gamma_4 / \gamma_0$ of dimension of velocity
characterizes hopping between a stacked atom and its the three unstacked
neighbors of its stacking partner. It also introduces difference between
the valence and conduction bands. To the leading order in $k$, this
hopping shifts the two middle bands (2 and 3) upward by a term
proportional to $v_4 k^2$ and shifts the two outer bands (1 and 4)
downward by the same amount. These effects of $\Delta$ and $v_4$ are
illustrated in Fig.~\ref{fig:BandAsym}.

Additional electron-hole asymmetry can in principle come from extrinsic
sources, e.g., charged impurities that can be present on or between the
layers. Besides creating a finite $V_\text{CN}$, these charges also move
$V = 0$ point away from the charge neutrality point $n = 0$. To the
first approximation,~\cite{McCann2006} this introduces an offset of the
interlayer bias: $V(n) \to V(n) + V_{0}$. However, our calculations
suggest that for reasonable $V_{0}$ this effect has a smaller influence on
the electron-hole asymmetry of the optical response than
$\Delta$ and $\gamma_4$.  

%%%%%%%%%%%%%%%%%%%%%%%%%%%%%%%%%%%%%%%%%%%%%%%%%%%%%%%%%%%%%%%%%%%%%%%%%%%%%%%%
%
% FIG. 6
%
%%%%%%%%%%%%%%%%%%%%%%%%%%%%%%%%%%%%%%%%%%%%%%%%%%%%%%%%%%%%%%%%%%%%%%%%%%%%%%%%

Based on the above discussion, we can predict qualitatively how the
position of the main conductivity peak should vary as a function of
$\delta V$. For example, on the electron side, and for $v_4 > 0$, the
peak should move to lower frequencies as $\delta V$ increases.
Alternatively, this can be seen from Fig.~\ref{fig:BandAsym}: the top
two bands move closer to each other as $k = k_F$ increases.

For the quantitative analysis, we use a full numerical calculation of $\sigma$
and $T$, which is discussed in Sec.~\ref{sec:Derivation} below. It demonstrates
that for the case of small $\Gamma$ the energy $E_3$ is indeed in a good
agreement with the computed peak position $\Omega_0$. However, the broadening
observed in experiments~\cite{Li2008, Wang2008, Kuzmenko2008} is appreciable, in
which case the formula $\Omega_0 = (E_2 + E_3) / 2$ is more accurate. Of course,
for fairly large $\Gamma$ other nearby transitions, $E_1$ and $E_4$, start to
influence the lineshape of the main peak. This is especially noticeable on the
hole side, where the $E_4$-peak is right next to the main one. In the
calculations this two-peak structure is unmistakable, see
Fig.~\ref{fig:sig_Omega}(b). In the experiment, where the main peak is for some
reason strongly enhanced compared to the calculation, the $E_4$ peak is somewhat
disguised. As pointed out by Kuzmenko,~\cite{Kuzmenko2008} the difference
between $E_4$ and $E_2$ can in principle provide a direct spectroscopic
measurement of the energy gap $V$.

For detailed comparison with experiment we use our numerical results rather than
Eqs.~\eqref{eqn:E2_asym} and \eqref{eqn:E3_asym}. Fitting them to the data, see
Fig.~\ref{fig:Tpeaks}, we have obtained estimates of $\gamma_1$, $\gamma_4$, and
$\Delta$ listed in Table~\ref{tbl:SWMc_values}. This fitting procedure proved to
be very straghtforward. For example, $\Delta$ is determined mostly by the
splitting of the peak positions on the electron and the holes sides of the
charge neutrality point. Parameter $\gamma_1$ is essentially the average of the
two. Finally, $\gamma_4$ controls the slope of the $\Omega_0(V_g)$ curves away
from $V_\text{CN}$. Therefore, all these parameters can be uniquely determined.

In Table~\ref{tbl:SWMc_values} we also list SWMc values suggested in
prior literature. They mainly agree with ours for the
principal SWMc parameters $\gamma_0$ and $\gamma_1$ but show some
deviations for the more subtle quantities $\gamma_4$ and $\Delta$ we
have been discussing here. Possible reasons for these differences are
given in Sec.~\ref{sec:Discussion}.
 
%%%%%%%%%%%%%%%%%%%%%%%%%%%%%%%%%%%%%%%%%%%%%%%%%%%%%%%%%%%%%%%%%%%%%%%%%%%%%%%%
\section{Derivation}\label{sec:Derivation}
\subsection{Band structure}

The bilayer is two monolayers stacked together, see Fig.~\ref{fig:band}(a). In
the bulk graphite the preferential stacking is the AB (Bernal) one, such that
only one sublattice of each layer is bonded to each other. In order to achieve
agreement with experiments,~\cite{Li2008} we have to assume that in the bilayer
the stacking is the same. We use the basis $\left\{ \Psi_{A1}, \Psi_{B1},
\Psi_{B2}, \Psi_{A2} \right\}$, where the letter stands for the sublattice label
and the number represents the layer index. In this basis the SWMc tight-binding
Hamiltonian for the bilayer becomes~\cite{Nilsson2008}
\begin{equation}
  \mathbf{H} =
  \begin{pmatrix}
    -\frac{V}{2} + \Delta & \phi & \gamma_1 & -v_4 \phi^{*} \\
    \phi^{*} & -\frac{V}{2} & -v_4 \phi^{*} & v_3 \phi \\
    \gamma_1 & -v_4 \phi & \frac{V}{2} + \Delta & \phi^{*} \\
    -v_4 \phi & v_3 \phi^{*} & \phi & \frac{V}{2}
  \end{pmatrix}\,,
  \label{eqn:bilayerH}
\end{equation}
where $\phi = -i (k_{x} + i k_{y})$ and $(k_x, k_y)$ is the deviation
of the quasimomentum from the $K$ point.

Given $V$, it is easy to obtain the four band energies $\varepsilon_\alpha(k)$
and the corresponding eigenstates $\left|\alpha, \mathbf{k} \right\rangle$
numerically. However, as mentioned in Sec.~\ref{sec:Results}, $V$ should be
determined self-consistently as a function of $V_g$, or equivalently, the total
carrier concentration $n$. The algorithm for doing so is given next.

\subsection{Electrostatics}
%%%%%%%%%%%%%%%%%%%%%%%%%%%%%%%%%%%%%%%%%%%%%%%%%%%%%%%%%%%%%%%%%%%%%%%%%%%%%%%%
%
% Fig. 7
%
\begin{figure}
  \centering
  \includegraphics[width=0.8\columnwidth]{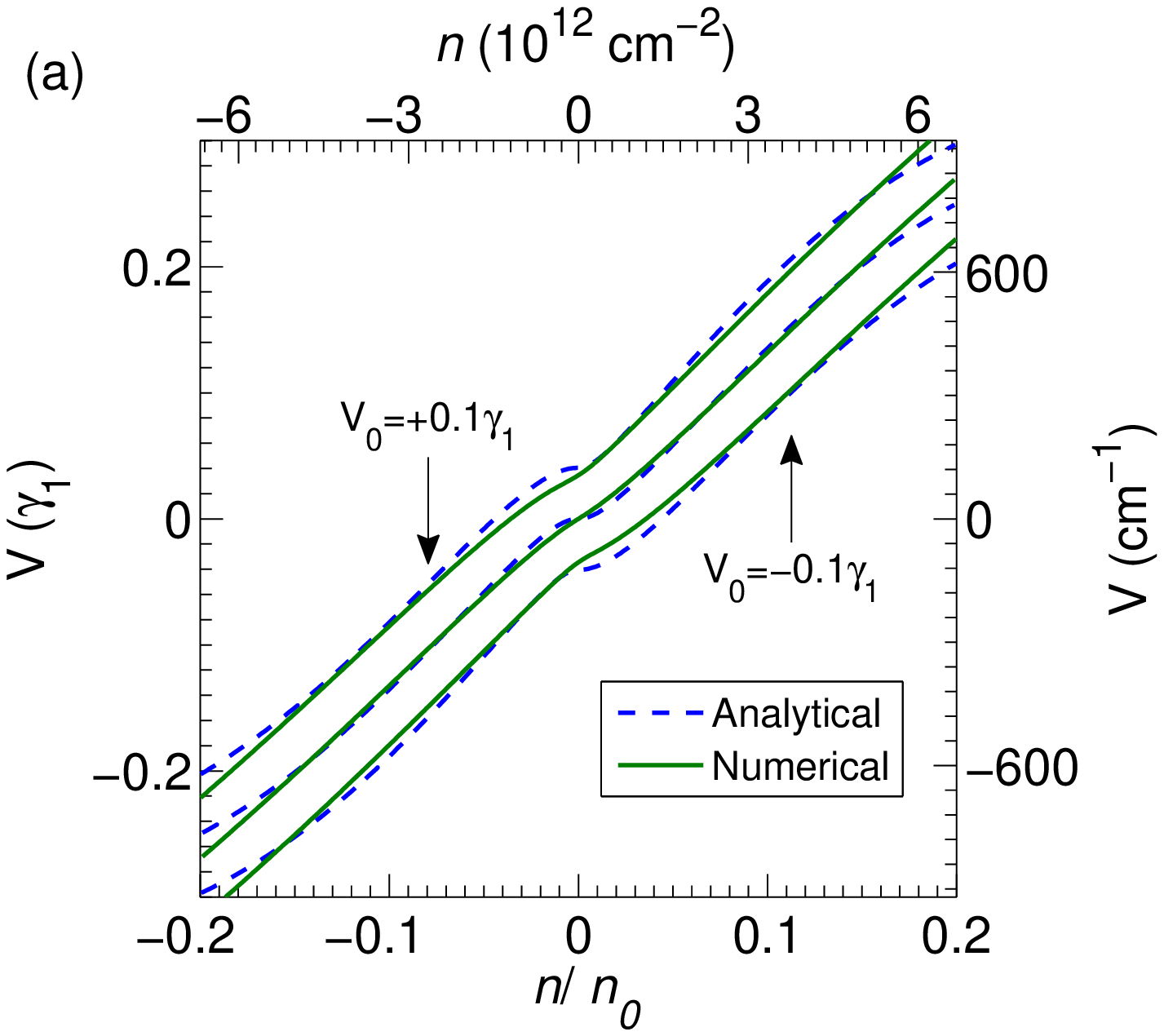} \\
  \includegraphics[width=0.8\columnwidth]{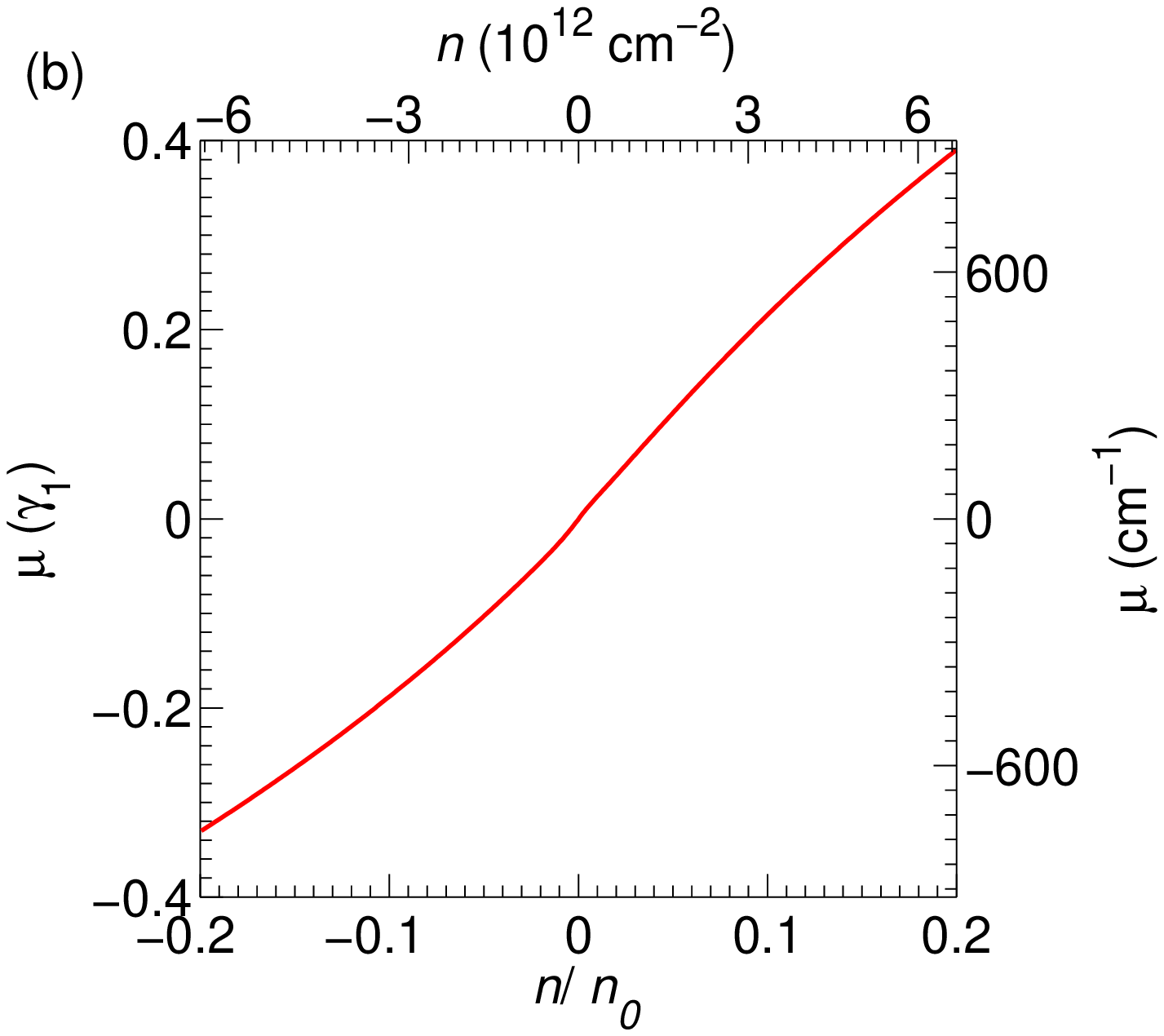}
\caption{(a) Interlayer bias $V$ as a function of total density $n$. Three sets
of curves correspond to (from top to bottom) $V_0 = 0.1\gamma_1$, $0$, and
$-0.1\gamma_1$. The dashed lines are computed from
Eq.~\eqref{eqn:McCann_broad}. (b) Chemical potential \textit{vs\/}. $n$ for $V_0
= -0.1\gamma_1$.\label{fig:electrostatics}
}
\end{figure}
%%
%%%%%%%%%%%%%%%%%%%%%%%%%%%%%%%%%%%%%%%%%%%%%%%%%%%%%%%%%%%%%%%%%%%%%%%%%%%%%%%%

As discussed in the literature,~\cite{McCann2006, Nilsson2008} the electric
field of the gate has two major effects on the bilayer graphene. First, it
modifies the bands by introducing a potential difference between the layers and
as a consequence opens up the energy gap. Second, it induces charge carriers.
Electric field of the charged impurities can play a similar role: it creates a
layer asymmetry $V_0$ and opens a gap at the charge neutral point much like an
external gate. But the more important effect of the impurities is presumably the
broadening of the electron energy states, which we describe by a
phenomenological constant $\Gamma$. For example, if the impurities are
distributed symmetrically between the two layers, then $V_0$ is zero but
$\Gamma$ is still finite. We assume $\Gamma$ to be real and independent of
energy, momentum, or a band index. This is certainly a very rudimentary
treatment of disorder compared to, e.g., self-consistent
schemes.~\cite{Ando2002,Nilsson2008,Stauber2008cos} However, since the source of
disorder in graphene is still debated, we think that this simple approach is
adequate for our purposes as long as $\Gamma$ is treated as another adjustable
parameter.

To compute $V(n)$ and $\mu(n)$ we set up a system of equations similar to those
in Refs.~\onlinecite{McCann2006} and \onlinecite{McCann2007}. These equations
capture the dominant Hartree term of the interaction but neglect exchange and
correlation energies.~\cite{Min2007} The first equation is
[cf.~Eq.~\eqref{eqn:n_from_V}]
\begin{equation} \label{eqn:n_total}
                n = n_{t} + n_{b} = C_b \delta V / e\,,
\end{equation}
where $n_{t}$ and $n_{b}$ are the carrier concentrations of the top and
bottom layers, and $C_{b}$ is the capacitance to the gate. Second, the
electrostatic potential difference between the two layers $V$ is given
by
\begin{equation}
  V = \frac{4\pi e^2}{\kappa} (n_t - n_b) c_{0}\,,
  \label{eqn:Delta}
\end{equation}
where $\kappa$ is the dielectric constant and $c_{0}$ is the distance
between the layers. Next, the Hamiltonian and hence the wavefunction and
the layer density $n_t$ and $n_b$ depend on $V$. Therefore the
quantities $V$, $n_t$, and $n_b$ must be solved for self-consistently. If
the broadening $\Gamma$ is neglected, this can be done analytically in
the limit $V, \mu \ll \gamma_1$, which gives
$V \simeq \mathcal{V}(n, V_0)$, where~\cite{McCann2006, McCann2007}
\begin{equation}
  \mathcal{V}(n, V_0) =
%  \frac{V_0 + (\pi n / n_0) \gamma_1}
%       {\Lambda^{-1} + (\pi |n| / n_0) - \frac{1}{2} \ln (\pi |n| / n_0)}\,,
  \frac{X \gamma_1 + V_0}
       {\Lambda^{-1} + |X| - \frac{1}{2} \ln |X|}\,,
\quad X = \frac{\pi n}{n_0}\,,
  \label{eqn:McCann}
\end{equation}
$n_0$ is defined by Eq.~(\ref{eq:n_0}), and $\Lambda\equiv e^2 c_0
n_0 / (\pi \kappa \gamma_1)$ is the dimensionless strength of the
interlayer screening. Using the typical parameter values, one
estimates~\cite{McCann2006} $\Lambda \sim 1$, and so
the interlayer screening is significant.~\cite{McCann2006, Min2007}

For experimentally relevant broadening $\Gamma \sim 0.02 \gamma_1$, the
approximation leading to Eq.~\eqref{eqn:McCann} is no longer accurate.
Therefore, we computed the dependence of $n_{t}$ and $n_{b}$ on $V$
numerically as follows. We first define the retarded Green's function
$\mathbf{G}^R$ by the analytic continuation $\mathbf{G}^R \left(
\varepsilon \right) = \mathbf{G}\left( \varepsilon\rightarrow\varepsilon
+ {i} \Gamma \right)$ of the following expression
\begin{equation} \label{eqn:Greensfunc}
  \mathbf{G}(\varepsilon) = \sum\limits_{\alpha = 1}^{4}
  \frac{1}{\varepsilon-\varepsilon_{\alpha}\left( k \right)}
  \left|\alpha, \mathbf{k} \right\rangle
  \left\langle \alpha, \mathbf{k} \right|\,.
\end{equation}
Then we compute $n_t$ from
\begin{equation}\label{eqn:n_t}
n_t = -\int \frac{d^2 k}{(2 \pi)^2}
       \int\limits_{-\infty}^{\mu} \frac{d \varepsilon}{\pi}
       \text{Im} [G^R_{11}(\mathbf{k}, \varepsilon)
                + G^R_{22}(\mathbf{k}, \varepsilon)]\,,
\end{equation}
using numerical quadrature. Similarly, the formula for $n_b$ is obtained by
replacing $G_{11} + G_{22}$ with $G_{33} + G_{44}$.

The system of nonlinear equations~(\ref{eqn:n_total}),
(\ref{eqn:Delta}), and (\ref{eqn:n_t}) is solved by an iterative
procedure. For a given chemical potential $\mu$ we start from some
initial guess on $V$. Then we diagonalize the Hamiltonian and compute
$\mathbf{G}^R$, $n_t$, and $n_b$. Substituting them into
Eq.~(\ref{eqn:Delta}), we get the value of $V$ for the next iteration.
(Actually, we use not this value directly but a certain linear
combination of the new and old $V$ to achieve convergence.) The
iterations terminate when the values of $V$ changes by less than a
desired relative accuracy (typically, $10^{-5}$). The results of these
calculations are in a good agreement with Eq.~\eqref{eqn:McCann} for
$\Gamma = 0$, and so are not shown. On the other hand, the results for
$\Gamma = 0.02 \gamma_1$, which are plotted in
Fig.~\ref{fig:electrostatics}, appreciably deviate from
Eq.~\eqref{eqn:McCann}. The agreement greatly improves (see
Fig.~\ref{fig:electrostatics}) if instead of Eq.~\eqref{eqn:McCann} we
use, on heuristic grounds, the following formula:
\begin{gather}
 V(n) = \mathcal{V}(n_*, V_0) - \mathcal{V}(n_\Gamma, 0)\,,
\label{eqn:McCann_broad}\\
  n_* = \text{sign}(n) \sqrt{n^2 + n_\Gamma^2}
\,\,,\quad
  n_\Gamma = \text{sign}(n) \frac{2\Gamma n_0}{\pi \gamma_1}\,.
\label{eqn:n_*}
\end{gather}
%%

%%%%%%%%%%%%%%%%%%%%%%%%%%%%%%%%%%%%%%%%%%%%%%%%%%%%%%%%%%%%%%%%%%%%%%%%%%%%%%%%
\subsection{Dynamical conductivity}

The above procedure enables us to compute $V$ and $n$ for a given
chemical potential $\mu$. With the former determining the Hamiltonian
and therefore its eigenstates, and the latter determining their
occupancy, we can now compute the dynamical conductivity by the Kubo
formula~\cite{Mahan1990}
\begin{equation}
  \sigma_{xx}\left(\Omega \right) = i\,
  \frac{\Pi^{R}_{xx}(\Omega) - \Pi^{R}_{xx}(0)}{\Omega + i 0}\,,
  \label{eqn:Kubo}
\end{equation}
where the polarization operator $\Pi^{R}_{xx}(\Omega)$ is given by
\begin{widetext}
  %% correlator in terms of Green's function
  \begin{equation}
      \Pi^{R}_{xx} (\Omega) =
      i g \frac{e^2}{\hbar^2} \int \frac{d^2 k}{(2\pi)^2}
      \int\limits_{-\infty}^{\mu} \frac{d\varepsilon}{2\pi}
      \text{Tr} \left\{
      \mathbf{v}_{x}
      \left[\mathbf{G}^{R}\left( \mathbf{k}, \varepsilon \right) -
      \mathbf{G}^{A}\left(\mathbf{k}, \varepsilon \right) \right]
      \mathbf{v}_{x}
      \left[ \mathbf{G}^{R}\left( \mathbf{k}, \varepsilon + \Omega \right)
      +
      \mathbf{G}^{A}\left( \mathbf{k}, \varepsilon - \Omega \right) \right]
      \right\}\,.
    \label{eqn:correlator}
  \end{equation}
In this equation $g = 4$ is the spin-valley degeneracy of graphene,
$\mathbf{v}_{x} = \hbar^{-1}
\partial \mathbf{H} / \partial {k_x}$ is the velocity operator,
and $\mathbf{G}^{R,A}$ at the
retarded and the advanced Green's functions. Assuming again that the broadening
is momentum and energy independent, these functions are obtained by
the analytic continuation of $\mathbf{G}$ in Eq.~\eqref{eqn:Greensfunc}:
$\mathbf{G}^{R,A}\left( \varepsilon \right) = \mathbf{G}\left(
\varepsilon\to\varepsilon\pm i\Gamma \right)$. After some algebra, we find
\begin{equation} \label{eqn:corr_2nd_form}
    \Pi^{R}_{xx} \left( \Omega \right) = {i} g \left( \frac{e}{\hbar} \right)^2
    \int \frac{d^2k}{\left( 2\pi \right)^2} \sum_{\alpha,\beta}
    \left| M_{\alpha\beta} (\mathbf{k}) \right|^2
    \sum_{\xi, \zeta = \pm 1} \xi
    K\left[ \varepsilon_{\beta}\left( k \right) - i \Gamma \xi,\;
    \varepsilon_{\alpha}\left( k \right) - \left( i \Gamma + \Omega \right) \zeta
    \right]\,,
\end{equation}
\end{widetext}
where $M_{\alpha\beta}\left( \mathbf{k} \right)=\left\langle \alpha,
\mathbf{k} \left| \mathbf{v}_x \right| \beta, \mathbf{k}
\right\rangle$ are the transition matrix elements and function
$K$ is defined by
\begin{equation} \label{eqn:Kernel}
    K(z_{1}, z_{2}) = \frac{\ln (\mu - z_1)  - \ln(\mu - z_2)}
    {2 \pi (z_1 - z_2)}
\end{equation}
with the branch cut for $\ln z$ taken to be $(-\infty, 0]$.

For vanishing $V$ and $\Gamma$ the conductivity can be computed in the closed
form, see Appendix~\ref{app:conductivity}. For other cases, we evaluated it
numerically. The results are shown in Figs.~\ref{fig:conductivity} and
\ref{fig:sig_Omega}. To demonstrate agreement with previous theoretical
calculations,~\cite{Abergel2007, Nilsson2008, Nicol2008, McCann2007} we present
$\sigma(\Omega)$ computed for a very small broadening $\Gamma$ in
Fig.~\ref{fig:conductivity}. In this case one can easily identify all six
transitions. As explained above, the sharp features at $\Omega\approx
3200\,\text{cm}^{-1}$ are due to the high optical density of states at energies
$E_2 < \hbar \Omega < E_3$. The other prominent feature at $\Omega = 0$ is the
intraband Drude peak. (Its height is related to the transport mobility.) In
Fig.~\ref{fig:sig_Omega} the calculation is done for much larger $\Gamma$ to
match the experimental data. This Figure has been discussed in detail in
Sec.~\ref{sec:Results}.

%%%%%%%%%%%%%%%%%%%%%%%%%%%%%%%%%%%%%%%%%%%%%%%%%%%%%%%%%%%%%%%%%%%%%%%%%%%%%%%%
\section{Discussion}\label{sec:Discussion}

In this paper we presented a joint experimental and theoretical study of the
infrared response of a bilayer graphene. Our results demonstrate a complex
interplay among various interband transitions and their disorder-induced
broadening. Nevertheless, by means of a careful analysis, we have been able to
explain the majority of the observed features within the conventional SWMc
model. The corresponding SWMc parameters are given in
Table~\ref{tbl:SWMc_values}, together with their estimated uncertainties. In
particular, our $\gamma_1$ should have a very high accuracy: about
$100\,\text{cm}^{-1}$, i.e., 3\%. The uncertainty in $\gamma_1$ comes
predominantly from an unknown systematic error that we make by neglecting the
renormalization of the spectrum by scattering processes. Since we assume that
the imaginary part $\Gamma \approx 65\,\text{cm}^{-1}$ of the electron
self-energy due to scattering is constant, its real part has to vanish by the
Kramers-Kr\"onig relations. In fact, this real part, which is generally
finite,~\cite{Nilsson2008} can shift the observed transition frequencies by an
amount that scales with $\Gamma$.

Let us now compare our SWMc parameters with those found in previous work on
bilayers and bulk graphite. For the bilayer case there is at present only one
other experimental determination~\cite{Malard2007} of $\gamma_j$'s. From
Table~\ref{tbl:SWMc_values} we see that the difference between our and their
values is primarily in $\gamma_1$. Actually, our SWMc parameters
can describe the Raman data equally well\cite{Nilsson_private} as those
given in Ref.~\onlinecite{Malard2007}. Our parameter values have smaller
estimated errors and should be considered more accurate.

In comparison with bulk graphite, the strongest discrepancy is in the value of
$\Delta$. The difference is significantly larger than the uncertainty of
$\Delta_\text{graphite}$ quoted in the early~\cite{Dillon1977,
Dresselhaus_graphite_review} and the recent experimental work,~\cite{Orlita2008}
which makes a strong case that $\Delta_\text{bilayer}$ differs from
$\Delta_\text{graphite}$ both in sign and in magnitude. To judge the true
significance of this result, one should recall that the physical meaning of
$\Delta_\text{bilayer}$ is the difference in the onsite energies of the A and B
sublattices.~\cite{Carter1953} However, in graphite the role of the same quantity is
played not by $\Delta$ but by the linear combination~\cite{Partoens2006}
\begin{equation} \label{eqn:Delta_prime}
   \Delta^\prime_\text{graphite} \equiv \Delta_\text{graphite}
                                  - \gamma_2 + \gamma_5\,. 
\end{equation}
For the sake of convenience, let us set $\gamma_2 = \gamma_5 = 0$ in the
bilayer, so that the A-B energy difference is equal to $\Delta^\prime$ in both
materials. Taking the most commonly used~\cite{Dresselhaus_graphite_review}
parameter values for graphite, we arrive at the remarkable empirical relation
\begin{equation} \label{eqn:Delta_prime_2}
\Delta^\prime_\text{graphite} \approx 37\,\text{meV}
\approx 2 \Delta^\prime_\text{bilayer}\,,
\end{equation}
which is much easier to interpret. Indeed, the physical origin of
$\Delta^\prime$ is the short-range (exponentially decaying with distance)
repulsion due to exchange and correlation effects between the electron states of
the stacked atoms. (Neither Coulomb nor even the van der Waals interaction have
short enough range to effectively discriminate between the two
sublattices,~\cite{Palser1999, Kolmogorov2005} given the relatively large
interlayer distance.) Since in the bilayer each A atom has a single stacking
partner while in the Bernal graphite it has two of them,
Eq.~\eqref{eqn:Delta_prime_2} is exactly what one would expect. More precisely,
it is expected if the interlayer distance in the bilayer and in the graphite are
nearly the same. The validity of Eq.~\eqref{eqn:Delta_prime_2} can be considered
an experimental evidence that this is indeed so.

%For this reason, we think that only the interactions among the nearest and the
%next-nearest neighbor layers are important for $\Delta$. (This is also in the
%spirit of the SWMc model.) Actually, in the Bernal (or AB) graphite the
%next-nearest-layer interactions cannot contribute to $\Delta$ simply by
%symmetry. This leaves us with the nearest-neighbor terms. Since $\Delta$ is
%numerically small, we can argue that it can be studied by the perturbation
%theory. In the bilayer each A atom has a single stacking partner but in the
%Bernal graphite it has two of them, and so naively we expect
%$\Delta_\text{graphite, AB} \approx 2 \Delta_\text{bilayer}$. In real graphite
%samples, which usually contain other stacking orders (rhombohedral and
%turbostratic), the ratio $\Delta_\text{graphite} / \Delta_\text{bilayer}$ can be
%less than $2$. This crude argument assumes of course that the interlayer
%distance in the bilayer and in the graphite are exactly the same, which may not
%be the case. Measurements of trilayers and tetralayers may shed further light on
%the question of the apparent difference of the SWMc parameters in few-layer and
%bulk graphite systems.

Another SWMc constant, which may seem to be different in the bilayer and the
bulk graphite is $\gamma_4$. As mentioned in Sec.~\ref{sec:Introduction}, this
is one of the parameters that in the past have been difficult to determine very
accurately. Our estimate of $\gamma_4$ can be defended on the grounds that (i)
it agrees with the Raman experiments~\cite{Malard2007} and (ii) it is
comparable to the accepted value of $\gamma_3$. These two parameters describe
hopping between pairs of atoms at equal distances in the lattice, see
Fig.~\ref{fig:band}(a), and theoretically are not expected to be vastly
different from each other. Large difference of $\gamma_4$ between the bilayer
and the bulk graphite is not expected either. Indeed, even when they disagree
about the order of magnitude (or sign) of $\Delta$, all electronic structure
calculations to date find that $\gamma_4 \sim \gamma_3$ and are of the same
order of magnitude in the two systems, see Table~\ref{tbl:SWMc_values}.

Parameter $\gamma_3$ itself cannot be reliably extracted from the experimental
data~\cite{Li2008} we analyzed here. At the relevant carrier concentrations the
main effect of $\gamma_3$ is to produce a weak trigonal warping of the band
dispersion.~\cite{Dresselhaus_graphite_review} This warping averages out over
the Fermi surface, and so has an effect similar to the broadening $\Gamma$: it
makes the $\gamma_1$ conductivity peak more symmetric and shifts it towards the
midpoint of $E_2$ and $E_3$, i.e., to slightly lower frequencies,
cf.~Figs.~\ref{fig:sig_Omega}(a) and (b). Thus, it is difficult to separate the
effect of $\gamma_3$ from the broadening due to disorder.

Regarding the latter, the dc mobility that we find from our numerically computed
$\sigma(0)$ using $\Gamma = 0.02\gamma_1 \approx 8\,\text{meV}$ is $\mu\approx
3900\,\text{cm}^{2}\,/\,\text{V}\text{s}$. This is close to the transport
mobility typical for bilayer graphene, supporting our interpretation that
$\Gamma$ arises mainly due to disorder.

Concluding the paper, we wish to draw attention to several features of the
experimental data that are not accounted for by our model. One of them is an
unexpectedly large amount of the optical weight in a range of frequencies below
the $\gamma_1$ peak. It is present between the Drude peak and $2 \mu$, i.e.,
twice the chemical potential. For the chosen $\Gamma$, our calculation predicts
$\text{Re}\, \sigma(\Omega) \sim 0.02 e^2 / \hbar$ at such $\Omega$, see
Fig.~\ref{fig:bilayeroptics}, whereas the measured value is several times
larger.~\cite{Li2008} This extra weight is present also in the monolayer
graphene, in the same range of frequencies.~\cite{Li2008a} A related issue is a
very gradual rise of $\text{Re}\, \sigma(\Omega)$ around the point $\Omega =
2\mu$ compared to a sharp threshold expected theoretically. These features can
be in part due to electron-phonon interaction~\cite{Stauber2008cos} or
midgap states~\cite{Stauber2008cos, Martin2008tci} but
other effects seem to be involved as well.

One very simple explanation would be to attribute both the broadening of the
$\Omega = 2\mu$ threshold and the extra weight at $\Omega < 2 \mu$ to long-range
density inhomogeneities in the sample. They can be caused by charge impurities
and remnants of the photoresist used in the sample processing. The presence of
such inhomogeneities would modulate the local chemical potential, and so in the
infrared response one would see a certain average of the $\sigma\left( \Omega
\right)$ taken at different $\delta V$. We illustrate this argument by
calculating the arithmetic mean of $\sigma\left( \Omega \right)$'s for positive
(negative) $\delta V$ and superimposing the results (shown by the dashed lines)
on the $\sigma\left( \Omega \right)$ traces for $\delta V = +50\,\text{V}$
($-50\,\text{V}$) in Fig.~\ref{fig:sig_Omega}(b). Such averaged conductivities
indeed resemble the experimental data [Fig.~\ref{fig:sig_Omega}(b)] more
faithfully. 

Another discrepancy between the experiment and the present theory is the
lineshape of the $\gamma_1$-peak. By varying $\Gamma$, we can fit either the
width or the height of the peak but not both. For example, in
Fig.~\ref{fig:sig_Omega}, where we chose to fit the width, the measured height of
the peak is sometimes nearly twice larger than the theory predicts. The extra optical
weight of the peaks appears to have been transferred from their high-frequency
sides, which are suppressed in experiment compared to the
calculations. These lineshape differences are significant enough to make us think that
some essential physics is still missing in the simple single-particle
picture presented in this paper. We speculate that including many-body effects
may be truly necessary for bringing theory and experiment to better agreement.

We are grateful to D.~Arovas, M.~Dresselhaus, A.~Kuzmenko, and K.~Novoselov for
illuminating discussions, to E. Henriksen, Z. Jiang, P. Kim, and H. L. Stormer
for providing the samples, experimental assistance, and fruitful discussions and
to A.~Castro Neto, V.~Fal'ko, and I.~Martin for valuable comments on the
manuscript. The work at UCSD is supported by the grants NSF DMR-0706654, DOE
DE-FG02-00ER45799, and by the UCSD ASC. The Advanced Light Source is supported
by the Director, Office of Science, Office of Basic Energy Sciences, under the
DOE Contract No. DE-AC02-05CH11231.

%%%%%%%%%%%%%%%%%%%%%%%%%%%%%%%%%%%%%%%%%%%%%%%%%%%%%%%%%%%%%%%%%%%%%%%%%%%%%%%%%%%%%%%%%%%%%
\appendix
\section{Reflection and Transmission}\label{app:R&T}

To compute the transmission coefficient $T$ and the reflection
coefficient $R$ we follow the standard procedure.~\cite{Jackson1998} In
general, the result depends on the angle of incidence and on the
polarization of light. Abergel and Fal'ko~\cite{Abergel2007a} derived
the formulas for $R$ and $T$ for the $S$-polarization where the electric
field is perpendicular to the plane of incidence (and parallel to the
sample surface). We reproduce them here with a slight change in
notation:
\begin{equation}
  \begin{aligned}
    R &= \left| -\frac
    {C\,n_{1} \cos\theta_1 - D \left[ \cos\theta_0 - 4\pi\sigma\right]}
    {C\,n_{1} \cos\theta_1 + D \left[ \cos\theta_0 + 4\pi\sigma \right]}
    \right|^2\,, \\
    T &= \left| -\frac
    {2\cos\theta_0\;n_{1}\cos\theta_{1}\;n_{2}\cos\theta_{2}}
    {C\,n_{1} \cos\theta_1 + D \left[ \cos\theta_0 + 4\pi\sigma \right]}
    \right|^2\,,
  \end{aligned}
  \label{eqn:RT_S}
\end{equation}
where $A$, $B$, $C$, and $D$ are given by
%%
%% definitions of A, B, C, and D
\begin{equation}
  \begin{aligned}
    A &= \cos\theta_2 \sin \delta_2
    + {i} n_2 \cos\theta_0 \cos \delta_2\,, \\
    B &= {i} \cos\theta_2 \cos \delta_2
    + n_2 \cos\theta_0 \sin \delta_2\,, \\
    C &= A\,n_1 \cos\theta_2 \sin \delta_1
    + {i} B\,n_2 \cos\theta_1 \cos \delta_1\,, \\
    D &= {i} A\,n_1 \cos\theta_2 \cos \delta_1
    + B\,n_2 \cos\theta_1 \sin \delta_1\,.
  \end{aligned}
  \label{eqn:ABCD}
\end{equation}
In Eqs.~\eqref{eqn:RT_S} and \eqref{eqn:ABCD}, the index $j=0,1,2$
represents vacuum, SiO${}_2$, and Si layers respectively, $n_j$ are the
index of refraction of each layer, and $\theta_j$ are the angles the light
ray makes with the surface normal in each layer. They satisfy Snell's law
$n_{j} \sin \theta_j = \text{const}$. Finally, $\delta_j = {k
L_j} / {n_j}$ is the phase
the light picks up as it makes one pass across the layer of thickness $L_j$.

For the other, $P$-polarization, where the electric field is not exactly
parallel to the surface of the sample, we find a different expression:
%%
%% R & T, P-polarization
\begin{equation}
  \begin{aligned}
    R &= \left| \frac
    {C\,n_{1} \cos\theta_0 -
    D\,\cos\theta_1 \left( 1-4\pi\sigma\cos\theta_0 \right)}
    {C\,n_{1} \cos\theta_0 +
    D\,\cos\theta_1 \left[ 1+4\pi\sigma\cos\theta_0 \right]}
    \right|^2\,, \\
    T &= \left| \frac
    {-2\cos\theta_0\,n_1\cos\theta_1\,n_2\cos\theta_2}
    {C\,n_{1} \cos\theta_0 +
    D\,\cos\theta_1 \left[ 1+4\pi\sigma\cos\theta_0 \right]}
    \right|^2\,.
  \end{aligned}
  \label{eqn:RT_P}
\end{equation}
For this polarization the conductivity enters $R$ and $T$ multiplied by
the cosine of the angle of incidence, i.e., its effect is reduced. In
our experiments, we typically have $\theta_0 \sim 30^\circ$,
and so this reduction is quite small. Its role is further diminished
by the presence of both polarizations in the infrared beam.
Thus, we decided not to include it in the analysis and do all the
calculation assuming the $S$-polarization only.

%%%%%%%%%%%%%%%%%%%%%%%%%%%%%%%%%%%%%%%%%%%%%%%%%%%%%%%%%%%%%%%%%%%%%%%%%%%%%%%%
\section{Conductivity of an unbiased bilayer at vanishing broadening}
\label{app:conductivity}

The conductivity for the case $\Gamma = V = 0$ was computed previously
in Refs.~\onlinecite{Abergel2007} and \onlinecite{Abergel2007a}.
In our attempt to reproduce their formula we discovered that
it contains a typographical sign error.~\cite{Comment_on_Abergel}
For future reference, we give the corrected expression below.

In the limit of zero broadening, $\Gamma\rightarrow 0$,
Eqs.~\eqref{eqn:Kubo}--\eqref{eqn:Kernel} reduce to
the following expression for the conductivity:
%%
%% bare conductivity
\begin{equation}  \label{eqn:sigma_bare}
    \sigma\left( \Omega \right) = \frac{g e^2 v^2}{2 i \pi \hbar}
    P\!\int_{0}^{\infty} \frac{d \omega}{\omega}
    \frac{ \Omega |M_{\alpha\beta}|^2}{\omega^2 - (\Omega + i 0)^2}
    \sum_{j} k_j(\omega) k_j^\prime(\omega)\,,
\end{equation}
where $P$ means principal value and the integration variable $\omega =
\left| \varepsilon_{\alpha} - \varepsilon_{\beta}\right|$ is the energy
difference between two states. The sum in Eq.~\eqref{eqn:sigma_bare} is
over all values of momentum $k_j(\omega)$ of which two states differing
in energy $\omega$ exist. For $V = 0$ where the
the matrix elements $M_{\alpha\beta}$ take a simple form, the
integration over $\omega$ in Eq.~\eqref{eqn:sigma_bare} can be done
analytically. The result can be written as a sum of three terms:
%%
%% Gamma = Delta = 0 analytic formula
\begin{equation}
  \frac{\sigma\left( \Omega \right)}{\sigma_0} =
  \tilde{\sigma}_{0}\left( \Omega \right) +
  \tilde{\sigma}_{\gamma_1}\left( \Omega \right) +
  \tilde{\sigma}_{2\gamma_1}\left( \Omega \right)\,,
  \label{eqn:sigma_analy}
\end{equation}
where $\sigma_0 = {e^2} / {\hbar}$ is the unit of conductivity,
$\tilde{\sigma}_{0}$ is contribution from transitions between bands 2
and 3 that turn on at $\Omega=0$, $\tilde{\sigma}_{\gamma_1}$ is
contribution from transitions between bands 1 and 3 and bands 2 and 4
that turn on at $\Omega=\gamma_1$, $\tilde{\sigma}_{2\gamma_1}$ is
contribution from transition between bands 1 and 4 that turn on at
$\Omega=2\gamma_1$. They are given by
\begin{subequations}
  \begin{align}
    \tilde{\sigma}_{0} &=
    \frac{g}{8}\left[ \frac{1}{2} \frac{\Omega+2\gamma_1}{\Omega+\gamma_1} -
    \frac{{i}}{\pi} \frac{\Omega \gamma_1}{\gamma_1^2-\Omega^2}
    \ln \left| \frac{\Omega}{\gamma_1} \right|\right] \,,\\
    \tilde{\sigma}_{\gamma_1} &=
    \frac{g}{8}\left[ \frac{\gamma_1^2}{\Omega^2}
    \Theta\left( \Omega -\gamma_1 \right)
    \!+\!\frac{{i}}{\pi} \left( \frac{2\gamma_1}{\Omega}
    %\right.\right. \notag\\ &\qquad-\left.\left.
    -\frac{\gamma_1^2}{\Omega^2}
    \ln \left| \frac{\gamma_1+\Omega}{\gamma_1-\Omega} \right| \right)
    \right]\,, \\
    \tilde{\sigma}_{2\gamma_1} &=
    \frac{g}{8}\left[
    \frac{1}{2}\frac{\Omega-2\gamma_1}{\Omega-\gamma_1}
    \Theta\left( \Omega -2\gamma_1 \right)
    \!-\!\frac{{i}}{\pi}\!\!\left(
    \frac{1}{2}\frac{\Omega^2-2\gamma_1^2}{\Omega^2-\gamma_1^2}
    \right. \right. \notag\\
    &\left.\left.
    \times \ln\left| \frac{2\gamma_1+\Omega}{2\gamma_1-\Omega} \right|
    \!+\!\frac{1}{2}\frac{\Omega \gamma_1 }{\Omega^2-\gamma_1^2}
    \ln\left| \frac{4\gamma_1^2-\Omega^2}{\gamma_1^2} \right|
    \right)\!\right]\,,
  \end{align}
\end{subequations}
where, for ease of notation, $\Omega$ stands for $\hbar \Omega$ and $g = 4$.

\bibliography{pubBilayerOptics}

\end{document}